\documentclass[12pt]{iopart}
\usepackage{tikz}
\usetikzlibrary{arrows.meta, positioning}
\usepackage{framed,color}
\definecolor{shadecolor}{rgb}{1,1,1}
\usetikzlibrary{decorations.pathreplacing}
\usetikzlibrary{decorations.markings}
\usetikzlibrary{arrows}

\usepackage[utf8]{inputenc}
\usepackage[T1]{fontenc}
\usepackage[english]{babel}
\usepackage{graphicx}
\usepackage{bm,bbm,amsthm,amsfonts,amssymb,amsopn,amsmath}
\usepackage{iopams,braket}
\usepackage{dsfont,mathtools,physics}
\usepackage{url,cite}
\usepackage{doi}
\usepackage{etoolbox}
\usepackage{enumitem}
\usepackage[makeroom]{cancel}
\usepackage{scalerel,color,pgf,tikz}
\usepackage{calc,xstring,ifthen}
\usepackage{pdftexcmds,adjustbox}

\usepackage[section]{placeins}

\newcommand{\be}{\begin{eqnarray}}
\newcommand{\ee}{\end{eqnarray}}

\definecolor{half}{rgb}{0.95,0.95,0.95}
\definecolor{full}{rgb}{0,0,0}
\definecolor{halfborder}{rgb}{0.8,0.8,0.8}
\definecolor{border}{rgb}{0.3,0.3,0.3}
\definecolor{colU}{rgb}{0.71,0.8,0.76}
\definecolor{colP}{rgb}{0.69,0.50,0.86}
\definecolor{colLines}{rgb}{0.31,0.31,0.31}
\definecolor{colObs}{rgb}{1,0.39,0.28}

\usetikzlibrary{decorations.pathmorphing,decorations.markings,patterns,decorations.pathreplacing,shapes.misc}

\newcommand\redemptyrectangle[2]{
  \draw[red,very thick] ({(#1)},{(#2-1)})  -- ({(#1+1)},{(#2)})  -- ({(#1)},{(#2+1)})  
  -- ({(#1-1)},{(#2)})  -- cycle;
}

\newcommand\redfullrectangle[2]{
  \draw[red,very thick,fill=full] ({(#1)},{(#2-1)})  -- ({(#1+1)},{(#2)})  -- ({(#1)},{(#2+1)})  
  -- ({(#1-1)},{(#2)})  -- cycle;
}

\newcommand\redundeterminedrectangle[2]{
  \draw[red,very thick,fill=halfborder] ({(#1)},{(#2-1)})  -- ({(#1+1)},{(#2)}) 
  -- ({(#1)},{(#2+1)}) -- ({(#1-1)},{(#2)})  -- cycle;
}

\newcommand\emptyrectangle[2]{
  \draw[border] ({(#1)},{(#2-1)})  -- ({(#1+1)},{(#2)})  -- ({(#1)},{(#2+1)})  
  -- ({(#1-1)},{(#2)})  -- cycle;
}

\newcommand\fullrectangle[2]{
  \draw[border,fill=full] ({(#1)},{(#2-1)})  -- ({(#1+1)},{(#2)})  -- ({(#1)},{(#2+1)})  
  -- ({(#1-1)},{(#2)})  -- cycle;
}

\newcommand\undeterminedrectangle[2]{
  \draw[border,fill=halfborder] ({(#1)},{(#2-1)})  -- ({(#1+1)},{(#2)}) 
  -- ({(#1)},{(#2+1)}) -- ({(#1-1)},{(#2)})  -- cycle;
}

\newcommand\rectangle[3]{
  \ifthenelse{\equal{#3}{1}}{\fullrectangle{#1}{#2}}{\ifthenelse{\equal{#3}{2}}{\undeterminedrectangle{#1}{#2}}
  {\emptyrectangle{#1}{#2}}};
}

\newcommand\redrectangle[3]{
  \ifthenelse{\equal{#3}{1}}{\redfullrectangle{#1}{#2}}{\ifthenelse{\equal{#3}{2}}{\redundeterminedrectangle{#1}{#2}}
  {\redemptyrectangle{#1}{#2}}};
}

\makeatletter
\def\@mkboth#1#2{}
\newlength\appendixwidth
\preto\appendix{\addtocontents{toc}{\protect\patchl@section}}
\newcommand{\patchl@section}{%
  \settowidth{\appendixwidth}{\textbf{Appendix }}%
  \addtolength{\appendixwidth}{1.5em}%
  \patchcmd{\l@section}{1.5em}{\appendixwidth}{}{\ddt}%
}
\makeatother
\usepackage{verbatim}
\usepackage{graphicx}
\usepackage{array}

\usepackage{amsmath}
\usepackage{tikz}
\usetikzlibrary{arrows.meta, positioning}
\usepackage{framed,color}
\definecolor{shadecolor}{rgb}{1,1,1}
\usetikzlibrary{decorations.pathreplacing}
\usetikzlibrary{decorations.markings}
\usetikzlibrary{arrows}
\usepackage{bm}
\usepackage{listings}
\usepackage[utf8]{inputenc}
\usepackage{amssymb}
\usepackage{amsfonts}
\usepackage{xcolor}
\definecolor{codegreen}{rgb}{0,0.6,0}
\definecolor{codegray}{rgb}{0.5,0.5,0.5}
\definecolor{codepurple}{rgb}{0.58,0,0.82}
\definecolor{backcolour}{rgb}{0.95,0.95,0.92}
\lstdefinestyle{mystyle}{
    backgroundcolor=\color{backcolour},   
    commentstyle=\color{codegreen},
    keywordstyle=\color{magenta},
    numberstyle=\tiny\color{codegray},
    stringstyle=\color{codepurple},
    basicstyle=\ttfamily\footnotesize,
    breakatwhitespace=false,         
    breaklines=true,                 
    captionpos=b,                    
    keepspaces=true,                 
    numbers=left,                    
    numbersep=5pt,                  
    showspaces=false,                
    showstringspaces=false,
    showtabs=false,                  
    tabsize=2}

\def\ms#1{{\color{blue} [MS: #1]}}
\begin{document}
\title[UniqueNESS]{UniqueNESS: Graph Theory Approach to the Uniqueness of Non-Equilibrium Stationary States of the Lindblad Master Equation}
\author{Martin Seltmann$^1$, Berislav Bu\v ca$^{2,}$$^{1,}$$^3$}
\address{$^1$Niels Bohr International Academy, Niels Bohr Institute, University of Copenhagen, Universitetsparken 5, 2100 Copenhagen, Denmark}
\address{$^2$Universit\'{e} Paris-Saclay, CNRS, LPTMS, 91405, Orsay, France}
\address{$^3$Clarendon Laboratory, University of Oxford, Parks Road, Oxford OX1 3PU, United Kingdom}
\eads{\mailto{martin.seltmann@nbi.ku.dk},
  \mailto{berislav.buca@universite-paris-saclay.fr}}
\begin{abstract}
The dimensionality of kernels for Lindbladian superoperators is of physical interest in various scenarios out of equilibrium, for example in mean-field methods for driven-dissipative spin lattice models that give rise to phase diagrams with a multitude of non-equilibrium stationary states in specific parameter regions. We show that known criteria established in the literature for unique fixpoints of the Lindblad master equation can be better treated in a graph-theoretic framework via a focus on the connectivity of directed graphs associated to the Hamiltonian and jump operators.
\\\\
Keywords: Quantum Dynamical Semigroups, Operator Algebras
\end{abstract}
\tableofcontents
\noindent
\section{Fixpoints in Quantum Dynamics}
\subsection{Fixpoint Sets}
The characterization of fixpoints (stationary states) for dynamical systems has a long history, starting with the study of classical mechanics via its intrinsic symplectomorphisms (canonical transformations) on phase space manifolds and the subsequent mathematical formalization in terms of fixpoint theorems; of particular importance to physics is the fixpoint theorem of Brouwer \cite{Brouwer1911} stating that any continuous map of a compact convex set to itself has an invariant point. The fixpoint theorem applies to all topologies and for most physical models the state spaces are indeed compact and convex.

In open quantum systems, the states are density operators $\rho$ that form a compact convex body within the space $\mathcal{B}(\mathcal{H})$ of bounded linear operators on a Hilbert space $\mathcal{H}$ and the set of fixpoints $\rho_\infty$ for any dynamics induced by a continuous superoperator mapping this convex shape to itself is thus guaranteed to be nonempty.

The study of quantum dynamics in many-body systems has challenged conventional thermalization paradigms and revealed a broad spectrum of nonequilibrium phenomena. In the past few decades, considerable theoretical, numerical, and experimental efforts have converged to build a comprehensive picture of how closed and open quantum systems evolve after external perturbations or sudden parameter changes.

Seminal reviews have established the scope and importance of nonequilibrium dynamics, outlining the role of quantum quenches, thermalization, and universal scaling laws\cite{Polkovnikov2011,Eisert2015,Dalessio2016} connect phenomena such as eigenstate thermalization with traditional statistical mechanics. These works laid the groundwork for understanding how isolated quantum systems evolve toward equilibrium despite their unitary dynamics.

The conceptual basis of nonequilibrium dynamics dates back to the discussion of thermalization in closed systems \cite{Deutsch1991,Srednicki1994} and has been expanded by studies showing that integrable systems often relax to nonthermal stationary states\cite{Rigol2007,Rigol2008}). In parallel, experimental advances in ultracold atoms and optical lattices have made it possible to directly study these dynamics \cite{Langen2015,Langen2015b}. Together, these studies set the stage for exploring both integrable and nonintegrable behavior within a unified framework.

The theoretical formulation relies heavily on many-body quantum field theory techniques; Keldysh diagrammatic approach\cite{Keldysh1965} and subsequent developments in non-equilibrium field theory \cite{Kamenev2011,Rammer2007} provide rigorous tools for analyzing time-dependent evolution. These methodologies are supported by insights from both semi-classical approximations and full quantum simulations, as described in \cite{Berges2007,Berges2004}.

Quantum quenches, where system parameters are suddenly altered, have become a paradigmatic setting to study non-equilibrium evolution. \cite{Calabrese2007,Calabrese2006} pioneered the application of conformal field theory to describe the spreading of correlations after a quench. This work was extended to the transverse-field Ising chain \cite{Calabrese2012}, revealing the emergence of light-cone-like dynamics, as later confirmed by \cite{Cheneau2012}.

Thermalization mechanisms in closed quantum systems have been intensely scrutinized. The eigenstate thermalization hypothesis (ETH), originally formulated in \cite{Deutsch1991,Srednicki1994}, explains how individual eigenstates of nonintegrable systems yield thermal expectation values. This phenomenon is revisited in integrable contexts by \cite{Rigol2007,Rigol2008} and further examined by \cite{Gogolin2016}. In addition \cite{Vidmar2016,Caux2012} have advanced the construction of generalized Gibbs ensembles for integrable systems, clarifying how conserved quantities control the dynamics.

Dynamical mean-field theories and time-dependent approaches have been essential in characterizing these phenomena. \cite{Moeckel2008} studied quenches in the Hubbard model, while \cite{Schiro2010} developed mean-field descriptions for correlated electron systems. These techniques complement more exact numerical approaches such as time-dependent density-matrix renormalization group (t-DMRG) methods \cite{White2004,Daley2004}.

Accurate simulation of non-equilibrium dynamics in many-body systems remains a formidable challenge. Numerical techniques like t-DMRG and exact diagonalization have been employed to track real-time evolution \cite{White2004,Daley2004}). Additionally, approaches based on time-dependent mean-field theory \cite{Schiro2010} and advanced quantum field theory methods \cite{Berges2008} have significantly contributed to our understanding.

The experimental investigation of non-equilibrium dynamics has been equally dynamic. Ultracold atom experiments \cite{Langen2015,Langen2015b} have allowed for controlled studies of quenches and thermalization, while Ramsey interferometry has been successfully used to measure spatiotemporal correlations in such systems \cite{Knap2013}. Moreover, experiments investigating the breakdown of adiabaticity in low-dimensional gapless systems \cite{Polkovnikov2008} highlight the interplay between system size, dimensionality, and dynamical response.

Despite significant progress, several questions persist in the study of non-equilibrium dynamics. Recent comprehensive reviews \cite{Mitra2018,Dziarmaga2010} stress the need for a deeper understanding of prethermalization phenomena \cite{Barmettler2009} and many-body localization \cite{Nandkishore2015}. At the same time, investigations into ultracold quantum gases \cite{Polkovnikov2011,Berges2010} have broadened the scope of non-equilibrium phenomena to include scenarios relevant for early universe cosmology.

Research on quenches in spin chains \cite{Essler2016} and studies of inhomogeneous quenches \cite{Sotiriadis2008} further illustrate the rich variety of behavior exhibited by non-equilibrium quantum systems. These diverse approaches converge to show that the interplay of interactions, integrability, and external driving leads to a landscape of dynamical behavior that is as complex as it is fascinating.

A crucial question is the one about ergodicity of quantum many-body dynamics. Examples of important ergodicity-breaking that have attracted much attention are many-body localization \cite{Nandkishore2015, lbits1,Abanin2019, Smith2016, Alet2018,DaleyMBL}, time crystals \cite{Wilczek2012, TCReview,Else2016, Else2020, Buca2019, Medenjak2020,VedikaLazarides,Sacha_2015,Sacha,Passarelli_2022,MeasurementTC,carollo2024quantum,DiehlTC,HadisehMulti,Shammah,Fabrizio1,Booker_2020,PhysRevA.108.L041303,PhysRevA.105.L040202,Wu_2024,Subhajit,PhysRevA.106.022209}, many-body scars \cite{Turner2018, Moudgalya2018, PhysRevB.101.165139,Schecter2019,Serbyn2021,chandran2022quantum,embed,HoshoScars,BridgingScars,Tom1,Leonardo1,Leonardo2,Leonardo3,Leonardo4,quasisymmetry}, fragmentation \cite{ZanardiFRAG,Fragmentation1,Fragmentation2,fragmentationSanjay,LOCfrag2,LOCfrag3,strictlylocalfrag,nicolau2023local,Zhang_2023,Sid,SanjayNew,Pozsgay,PhysRevB.107.205112,ArnabFragmentation,openFRAG}, and related forms of ergodicity breaking (e.g. \cite{Kormos_2016,Majidy_2024,CatQUBIT,Henrik,Olalla1,Olalla2,Kollath1,ETH1,PhysRevLett.134.073604,Buca2022Syn,Hosho,Jad4,Claeys2022emergentquantum,FiniteFreqDrude,AndreasStark1,PhysRevResearch.7.013178,PhysRevA.105.L020401,Shovan,PhysRevLett.132.020401,marche2025mathrmsu3fermihubbardgasthreebody}).

In summary, the study of non-equilibrium quantum dynamics has evolved into a vibrant field characterized by interdisciplinary approaches and diverse experimental platforms. Foundational works \cite{Polkovnikov2011,Eisert2015,Dalessio2016} have laid out the basic principles and techniques, while subsequent studies have refined our understanding of thermalization \cite{Rigol2008,Gogolin2016} and phase transitions \cite{Heyl2013,Jurcevic2017}. Cutting-edge computational methods \cite{White2004,Daley2004}) and innovative experiments \cite{Langen2015,Knap2013} continue to drive the field forward.

As the field moves into the future, the integration of methods from analytic theory \cite{Cazalilla2006,Calabrese2007} to advanced numerical simulations \cite{Barmettler2009,Essler2016}) ensures that non-equilibrium quantum dynamics will remain at the forefront of fundamental physics. The continuing dialogue between theory and experiment, as highlighted in works across the spectrum \cite{Moeckel2008,Bukov2015,Polkovnikov2008}, promises to unlock deeper mysteries regarding the emergence of statistical mechanics from quantum dynamics.

Finally, the synthesis of these diverse studies not only enhances our understanding of specific models from the Luttinger liquid (Cazalilla and colleagues \cite{Cazalilla2007,Iucci2009}) to the Hubbard model \cite{Moeckel2008}, but also informs broader questions regarding equilibration and universality in quantum systems \cite{Dziarmaga2010,Mitra2018}. With continuing advances in both theory and experimental control, nonequilibrium quantum dynamics remains a rich and rapidly developing arena for future research \cite{Berges2008,Berges2010}.

Related to questions of ergodicity is the question of when does the quantum Liouvillian display long-time dynamics, e.g. \cite{Buca2022Syn,Buca2023} and multiple fixpoints \cite{BucaProsen,AlbertJiang,BaumgartnerNarnhofer}

The kernel of the Liouvillian $\mathcal{L}$ is the space of all fixpoints $\rho_\infty$ of the dynamics, usually called non-equilibrium stationary states (NESS) due to the general possibility of couplings to various reservoirs with discrepancies in their thermodynamic potentials.

While specially tailored Liouvillians $\mathcal{L}$ with multiple steady states can certainly be constructed via reservoir engineering, they constitute a set of measure zero within the realm of all possible $\mathcal{L}$ and generic systems in finite dimensions possess a single fixpoint as the unique long-time limit of the Lindblad equation. The question about the dimensionality of the kernel is nevertheless of physical interest, as meanfield methods for driven-dissipative lattice models starting with products of density matrices on individual sites as approximations for the multi-site density matrix often result in the coexistence of multiple NESS in certain parameter regions of the phase diagrams.
\subsection{Semigroup Generators}
The dynamics is given by a time evolution superoperator $\tau_t$ acting on the observable algebra $\mathcal{A}\subset\mathcal{B}(\mathcal{H})$, evolving any operator $A\in\mathcal{A}$ in the Heisenberg picture via 
\begin{equation}
\tau_t(A)=\exp(+itH)A\exp(-itH)=:A(t)
\end{equation}
with Hamiltonian $H$ for the total closed system, while open system dynamics $\tau_t(A)$ is no longer described by conjugation with $U_t=\exp(itH)$, but via more general operator mappings: for any invertible superoperator $\mathcal{E}: \mathcal{B}(\mathcal{H})\rightarrow \mathcal{B}(\mathcal{H})$ preserving hermiticity, its inverse and derivatives (in case of a parametrized family of superoperators) are equally hermiticity preservation (HP) maps. Thus the composition $\mathcal{E}^{-1}\circ \dot {\mathcal{E}}=:\mathcal{L}$ of first derivative $\dot {\mathcal{E}}$ and left inverse $\mathcal{E}^{-1}$ can be decomposed as \begin{equation}\mathcal{L}(\cdot)=\sum\nolimits_l \gamma_l L_l(\cdot) {L_l}^*\end{equation} with $\gamma_l \in \mathbb{R}$, as first shown by de Pillis \cite{dePillis1967} for any linear HP map. The further restriction to trace erasure (TE) entails unit erasure (UE) of the dual, namely \begin{equation}\mathcal{L}^{*}(I)=\sum\nolimits_l \gamma_l {L_l}^*{L_l}=0\end{equation} and therefore
\begin{equation}
\mathcal{L}(\cdot)=\sum_l \gamma_l \left[L_l(\cdot) {L_l}^* - \{{L_l}^* L_l,\cdot\}/2\right] = \sum_l \gamma_l\left\{[L_l,\cdot {L_l}^*] + [L_l\cdot,{L_l}^*]\right\}/2
\end{equation}
for any HPTE superoperator. Trace erasure of a generator corresponds to trace preservation (TP) of the exponentiated map and thus all such $\mathcal{L}$ generate HPTP maps via exponentiation $\exp(\mathcal{L})$. Restricting further to superoperators with complete positivity (CP) and a semigroup structure
\begin{equation}
	\tau_0=I \quad \tau_t \circ\tau_s= \tau_{s+t} \quad s,t\in \mathbb{R}_+
\end{equation}
of the CPTP time propagators, Kossakowski with his collaborators Gorini and Sudarshan \cite{Gorini1976,fazio2025manybodyopenquantumsystems} achieved in a finite-dimensional setting for $\mathcal{B}(\mathcal{H})\simeq\mathbb{C}^{N\times N}$ the full characterization
\begin{equation}
\mathcal{D}(\cdot):=\sum_{k,l = 1}^{N^2-1}c_{kl}\left\{[F_k,\cdot F_l^*]+[F_k\cdot,F_l^*]\right\}/2
\end{equation}
as the dissipative part of the \textit{GKS-generator} $\mathcal{L}(\cdot)=-i[H,\cdot]+\mathcal{D}(\cdot)$ for completely positive semigroups with the \textit{Kossakowski matrix} $C:=(c_{ij})\geq 0$ and a traceless orthonormal basis $\{F_k\}$ except for the tracefull identity $I_N=:F_{N^2}$ taken as last basis element to span all $N^2$ dimensions of the matrix space and an effective traceless Hamiltonian $H=H^*$ that in general includes Lamb shifts from renormalization effects due to the environment.
\\\\
Simultaneously and in infinite dimensions using different techniques for the algebra of bounded operators on (separable) Hilbert spaces, Lindblad \cite{Lindblad1976} arrived at the same generator in its diagonalized form
\begin{equation}
\begin{split}
\mathcal{D}(\cdot) &=\sum\nolimits_l\left\{[L_l,\cdot L_l^*] + [L_l\cdot,L_l^*]\right\}/2\\ &=\sum\nolimits_l L_l\cdot L_l^* - \{L_l^* L_l,\cdot\}/2\\ &= \Phi(\cdot) - \{\Phi^*(I),\cdot\}/2
\end{split}
\end{equation}
that in the last line has been rewritten as its modern (standard) form using a CP map $\Phi$ (and its dual $\Phi^*$) with Kraus operators $L_l$.
The standard Lindblad equation can easily be obtained from the GKS-generator by diagonalization of the positive matrix $C=UDU^{-1}$ with transformation matrix $U=(u_{ij})$ as well as redefinition of the basis via $\widetilde{L}_l:=\sum_{l,m=1}^{N^2-1} u_{lm} F_m$ and absorbing the eigenvalues $\lambda_l$ in $D=\text{diag}(\cdots \lambda_l \cdots)$ into the \textit{Lindblad operators} $L_l:=\sqrt{\lambda_l} \widetilde{L}_l$ that give the respective Kraus decomposition of $\Phi$.
\\\\
The non-uniqueness of Kraus representations (all related via unitaries) for the CP map $\Phi$ leads to an invariance of $\mathcal{D}$ in diagonal form under unitary transformations
\begin{equation}
	L_l\rightarrow \sum\nolimits_m u_{lm}L_m
\end{equation}
among the Lindblad operators. Moreover, the full Lindbladian
\begin{equation}
\mathcal{L}(\cdot)=-i[H,\cdot]+\sum\nolimits_l L_l(\cdot) L_l^* - \{L_l^* L_l,\cdot\}/2
\end{equation}
is invariant under the "gauge" symmetry
\begin{equation}
\begin{split}
L_l&\rightarrow L_l+\alpha_k I\\ H&\rightarrow H-i\sum\nolimits_k (\alpha^*_k L_k-\alpha_k L_k^*)/2+\beta I
\end{split}
\end{equation}
with arbitrary scalars $\alpha_k\in\mathbb{C}$ and $\beta\in\mathbb{R}$, where $\beta I$ merely constitutes an irrelevant global energy shift.
\\\\
The division between coherent and incoherent parts is thus highly arbitrary, as the gauge freedom allows transformations between them. The separation is however unambigious for the generator in GKS-form: here the zero trace of its Hamiltonian \(H\) is enforced and any fixed traceless basis \(\{F_k\} \) determines a unique \(H\), so that for GKS-generators it makes sense to speak about processes like pure dissipation. Note that after diagonalization of the Kossakowski matrix and redefinition of the basis, the obtained Lindblad basis remains orthonormal and traceless - but the gauge transformations destroy both properties. The apparent simplicity gained with the standard form is therefore counterbalanced by the loss of orthonormality and tracelessness of general \(L_l\), unless explicitly enforced via the respective Kraus decomposition of \(\Phi\).
\subsection{Existing Algebraic Criteria}
Criteria for the existence of unique steady states have been established in the literature \cite{Spohn1976, Spohn1977, Evans1977, Frigerio1977, Frigerio1978,Zhang_2024,umanita2006classification,SchirmerWang} concerning properties of either the Kossakowski matrix $C$ or the Lindblad operators $L_i$ (Spohn 76/77, Evans 77, Frigerio 77/78) and NESS uniqueness is in particular guaranteed for the following three cases:
\\\\
1. Sufficient Condition on Kossakowski Kernel and Hilbert Space
$$\boxed{\dim(\ker(C))<\dim(\mathcal{H})/2}$$
2. Sufficient Condition on Lindblad Bicommutant and Operator Algebra
$$\boxed{
\{L_i,L_i^*\}''= \mathcal{B}(\mathcal{H})} $$
3. Necessary and Sufficient Condition on Extended Lindblad Commutant
$$\boxed{
\{H,L_i,L_i^*\}'= \mathbb{C} I} $$
This last criterion applies only for the existence of a faithful (full-rank) steady state. With the von Neumann bicommutant theorem for $W^*$-algebras, it amounts to the following useful statement: in case of faithfulness, the system has a unique NESS $\rho_\infty$ iff the set $\{H,L_i,L_i^*\}$ generates the full operator algebra $\mathcal{B}(\mathcal{H})$.
\\\\
Recently another sufficient uniqueness condition \cite{Yoshida2024} that additionally ensures the strict positivity of $\rho_\infty$ has been identified: if the set \begin{equation}\{H-i\sum_n L^*_nL_n/2,L_i\}\quad\text{(Yoshida Criterion)}\end{equation} generates the full operator algebra $ \mathcal{B}(\mathcal{H})$, then there exists a unique NESS $\rho_\infty>0$. In the special case of only self-adjoint $L_i$, this unique positive (faithful, full-rank) NESS is entirely specified as $\rho_\infty=I/\dim(\mathcal{H}).$
\subsection{Motivation for New Method}
While the latter criterion for a unique positive-definite NESS was proven in a simple way via elementary linear algebra, the general check for the validity of the condition involves more advanced mathematical techniques including representation and graph theory.\\\\ It should be noted from the outset that much of the literature on the generation of algebras automatically presumes unitality of the generator set, meaning that most proofs are based on the implicit assumption of a unit element among the given generators; this (often hidden) prerequisite rules out an application of all respective theorems to our case without prior modifications.
\\\\
Several authors have conveniently focused on the Lie algebra generated by the given operators via commutation, sometimes already yielding the desired operator algebra. This is successful whenever the structure constants of an algebra are entirely antisymmetric and thus only the Lie part is relevant; splitting the multiplicative operation for two algebra elements $a$ and $b$ via
\begin{equation}ab = (ab+ba)/2  + (ab-ba)/2 = \{a,b\}/2 + [a,b]/2
\end{equation}
into its symmetric and antisymmetric part leads to an analogous separation of the structure constants, as is quite familiar for the Pauli spin operators
\begin{equation}\sigma_a\sigma_b=\{\sigma_a,\sigma_b\}/2 + [\sigma_a,\sigma_b]/2 = \delta_{ab}I+i\varepsilon_{abc}\sigma_c
\end{equation}with symmetric Kronecker delta $\delta_{ab}$ and antisymmetric Levi-Civita epsilon $\varepsilon_{abc}$ as structure constants. This example illustrates the crucial role of the mentioned unitality assumption: the antisymmetric Lie part alone yields the third spin operator $\sigma_c$ for two given spin operators $\sigma_a$ and $\sigma_b$, but the unit element $I$ is never reached via commutation and the full spin algebra cannot be obtained with commutators unless the identity is naturally included in the set of generators. The single-spin case and the impossibility to reach $I=\sigma_a^2=\sigma_b^2=\sigma_c^2$ in the Lie sector generalizes to any non-projective operator: nested commutators cannot produce the square of a given generator $a\neq a^2$ and the symmetric part $\{a,a\}/2$ is necessary for this new element.           
\section{Graph Theory Method}
\subsection{Derivation \& Proof}
As shown in the above section, it is in general not sufficient to consider only the Lie part of an operator algebra: as the antisymmetric components do not necessarily reproduce the algebra, Jordan sectors are indispensable and the full reach of a given set of generators has to be studied. For an algebra $A$ of linear transformations on a linear space $V$, this notion of reachability is captured by the property of transitivity: $A$ is transitive iff $\{a|v\rangle| a \in A\} = V$ for every (nonzero) element $|v\rangle \in V$. The first observation is that in finite dimensions transitivity of any operator algebra implies that it is in fact the full operator algebra.
\\\\
\textbf{Claim: Relation Transitivity $\leftrightarrow$ Completeness}\\
In a finite-dimensional setting with $1 < \dim(V) < \infty$, any transitive algebra $A$ of linear transformations on a linear space $V$ is the algebra of all linear transformations on $V$.
\\\\
\textbf{Proof}\\
For an operator algebra $A$ to be transitive, it has to be unital (otherwise $\{a|v\rangle| a \in A\}$ would not include the vector $|v\rangle=I|v\rangle$ and thus not equal $V$) and therefore allows for a spectrum 
\begin{equation} \sigma(a):=\{z\in\mathbb{C}|a-zI\text{ noninvertible}\}\end{equation} defined for any $a \in A$. In addition to the unit element, the algebra $A$ must also contain a nontrivial noninvertible operator $n\neq 0$: should any nondiagonal operator $o$ be invertible, $n := o(sI-o)$ with $s\in \sigma(o)$ is certainly noninvertible. Its image $n(V)$ is therefore lower-dimensional than the space $V$; the restriction of the operator algebra $A$ to this lower dimension is of course still transitive on $n(V)$ and the claim can now be proven by induction in the dimension of the space the operators act on.
\\\\
\textbf{Dimensional Induction}\\
The base case $\dim(V)=1$ is trivially true and the induction step from lower dimensions to $\dim(V)$ relies on the induction hypothesis that the transitivity of $A$ restricted to $n(V)$ implies that all operators (linear transformations of any rank) on $n(V)$ are contained in this algebra restriction - in particular the existence of an operator $r\in A$ with 
$nrn$ being a transformation of rank one on $V$, which is therefore some outer product: $nrn=|f\rangle\!\langle g|$ in Dirac notation. As the algebras $A$ and dual $A^*$ are assumed to be transitive, all operators of rank one are reached via $a(|f\rangle\!\langle g|)$ for $a\in A$ together with $(|f\rangle\!\langle g|)a$ for $a\in A^*$ and thus (all operators being sums of these building blocks) the full operator algebra.\\\\
The fact that the transitive property is characterized by all generated spaces $\{a|v\rangle| a \in A\}$ entails the equivalence of transitivity and irreducibility for an operator algebra $A$: it is irreducible iff there exist no nontrivial invariant subspaces, which is the case iff all subspaces $\{a|v\rangle| a \in A\}$ for any nonzero element $|v\rangle \in V$ equal the total linear space $V$.
\\\\
On the whole, the above line of reasoning has established identical notions for a generated algebra of operators to be the maximal possible one (full algebra) via the triple 
\begin{equation}\text{Irreducibility} \Leftrightarrow \text{Transitivity} \Leftrightarrow \text{Maximality}\end{equation}
valid for any operator algebra in finite dimensions greater than one over a field with algebraic closure; so a check for generation of an entire algebra is in fact equivalent to a check for reducibility. In other words the complete algebra is fully distinguished by lack of nontrivial invariant subspaces (irreducibility) and lack of nontrivial irreducible subalgebras (uniqueness).
\\\\
Our check for (ir)reducibility of an operator algebra obtained from a set $S$ of generators now boils down to an investigation of subspace invariances for every generator $g \in S$; the matrix representation of $g$ is irreducible iff it does not map any nontrivial subspace into itself.
\\\\
\textbf{Algebraic Graph Theory}\\
The mathematical discipline of graph theory \cite{Diestel2017} provides an efficient and direct method to examine matrix reducibility: the main idea is that all matrices $g$ are interpreted as \textit{adjacency matrices} of an associated directed graph (digraph) defined in the following way: the digraph $D(g)$ is a set $X$ of $v:=\dim(V)$ elements called vertices together with a set $E\subset X\!\times\! X$ of ordered vertex pairs called edges such that $(x_i,x_j)\in E$ whenever $g_{ij}\neq 0$, the corresponding edge connects $x_i$ with $x_j$ in the directed manner of an arrow pointing from the former to the latter. 
\\\\
\textbf{Connectivity}\\
Strong connectivity between vertices is the existence of a path connecting them in both directions (mutual reachability) and defines an equivalence relation on $X$, partitioning the digraph into strongly connected components such that vertices from the same component are mutually reachable from each other and vertices from different components are not mutually reachable.
\\\\
\textbf{Claim: Relation Reducibility $\leftrightarrow$ Connectivity}\\
The matrix $g$ is irreducible iff the associated digraph $D(g)$ is one strongly connected component. 
\\\\
\textbf{Proof}\\
Any reducible $g$ entails at least one nontrivial invariant subspace of $V$ and must therefore be similar to a block-triangular matrix; the respective permutations of rows and columns do not alter the connectivity of the associated digraph, as this similarity transformation does not change the underlying graph structure and merely relabels its vertices. But a block-triangular form of the matrix implies that the digraph disintegrates into two components without a bidirectional connection between them, hence $D(g)$ is not strongly connected. Conversely, any graph without strong connectivity allows for the partitioning of the vertex set into components that are without a bidirectional connection and give rise to a block-triangular structure of the corresponding adjacency matrix that is thus reducible.
\\\\As a consequence, there is a clear correspondence between the connectivity of a digraph and the reducibility of its adjacency matrix. In fact, utilizing the relation between matrices and graphs gives valuable insight into the pattern of stable subspaces: decomposing the full linear space $V$ into the direct sum $V=V_1 \oplus V_1^\perp$ of the smallest invariant subspace $V_1$ and its complement $V_1^\perp$, the matrix representation $g$ of the operator according to this split is a block-triangular one; repeating this process for $V_1^\perp$ recursively with a chain of $k$ invariant subspaces $V_i$ in ascending order of dimension leads to the decomposition
\begin{equation} V=\bigoplus_i V_i  
\end{equation}
and the matrix shape
\begin{equation} g = \begin{bmatrix}g_1&g_{1,2}&\ldots&g_{1,k}\\0&g_2&\ldots&g_{2,k}\\\vdots&\vdots&\ddots&\vdots\\0&0&\ldots&g_k\end{bmatrix}\end{equation}
with irreducible diagonal submatrices $g_i$. The digraphs $D(g_i)$ for these matrix blocks are strongly connected and called the strong components of the total digraph $D(g)$, while nonzero nondiagonal blocks $g_{i,j}$  correspond to the (unidirectional) connections between these strong components; the lack of paths in the other direction is reflected by the respective zero blocks mirrored along the diagonal. 
\\\\
In the same way every matrix algebra can be put into block-triangular form with diagonal blocks representing irreducible subalgebras: with the above procedure, any  subalgebra $A\subset L(V)$ of linear transformations on $V=\oplus_i V_i$ is compartmentalized as
\begin{equation}L(V)\supset A = \begin{bmatrix}L(V_1)&\ast&\ldots&\ast\\0&L(V_2)&\ldots&\ast\\\vdots&\vdots&\ddots&\vdots\\0&0&\ldots&L(V_k)\end{bmatrix}\end{equation}
where $L(V_i)$ denotes the irreducible subalgebra on $V_i$ and therefore the full algebra of linear transformations on $V_i$ according to the equivalence demonstrated in this chapter.
\\\\
A digraph $D(S)$ can now be assigned to the generator set $S$ of an algebra $A$ with a directed edge $(x_i,x_j)\in E$ whenever there exists an 
$a\in S$ with matrix entry $a_{ij}\neq 0$. If the generated algebra $A$ is in fact $L(V)$, then $D(S)$ is strongly connected - a property that can be efficiently checked via computer algebra systems in linear time, meaning the automated check has time complexity $O(|X|+|E|)$.
\\\\
We showcase this method for the driven-dissipative model with spin qubits on $N$ sites of hypercubic lattices in $D$ spatial dimensions undergoing both flip-flop interactions and Rabi oscillations via the Hamiltonian in the rotating (drive) frame
$$
H=\sum_{\left\langle i, j\right\rangle} J\left(\sigma_i^{+} \sigma_j^{-}+ \sigma_j^{+} \sigma_i^{-}\right)+ \sum_i\left(\Delta_z \sigma_i^{z}+\Delta_x \sigma_i^{x}\right)
$$
with hopping amplitude $J$ and spin operators $\{\sigma_i^k|k\in \{x,y,z\}\}$ for each site $i$, Rabi frequency $\Delta_x$ and detuning $\Delta_z$ between oscillatory driving field and resonant transition frequencies as well as dissipation via jump operators $L_i = \sqrt{\gamma_i} \sigma_i^{-}$ and thus the corresponding Lindbladian dissipator 
$$
\mathcal{D}(\cdot)=\sum_i \gamma_i\left(\sigma_i^{-} (\cdot) \sigma_i^{+}-\left\{\sigma_i^{+} \sigma_i^{-},\cdot\right\}/2\right)
$$
for spin loss on each site with mutual independence and individual rates $\gamma_i$. This spin lattice model exhibits bistability in numerical computations even though the Lindbladian superoperator yields an exclusive NESS $\rho_\infty$ in analytic calculations.
\\\\
While the third (necessary and suffient) uniqueness condition is met (as $\sigma_i^{+}$ and $\sigma_i^{-}$ together generate the spin operator algebra on each site and therefore the set $\{L_i,L_i^*\}$ of all Lindblad operators and their adjoints generates the full algebra irrespective of $H$), the Yoshida criterion for the uniqueness of a strictly positive $\rho_\infty$ concerns the algebra generated by the set
$$\left\{
\sum_{\left\langle i, j\right\rangle} J\left(\sigma_i^{+} \sigma_j^{-}+ \sigma_j^{+} \sigma_i^{-}\right)+ \sum_i\left(\Delta_z \sigma_i^{z}+\Delta_x \sigma_i^{x}\right)-i\sum_i \sigma_i^{+}\sigma_i^{-}/2, \sqrt{\gamma_i}\sigma_i^{-} \right\}$$
and is verified in Python via an efficient ``connectivity check" code.
\subsection{Generated Algebras}
To develop an intuitive understanding of the possible patterns that arise for different generators, it is helpful to look at the algebra produced by a generator. In its generalized eigenbasis, a generator $g$ assumes the Jordan normal form $J=\oplus_i  J_{\lambda_i,m_i}$ with Jordan blocks $J_{\lambda_i,m_i}$ of dimension $m_i$ for each eigenvalue $\lambda_i$ respecting geometric multiplicities.
\\\\
The algebra generated by $J$ is in fact the polynomial algebra $\mathbb{C}[J]$ given by all expressions
\begin{equation}
p(J)=\sum_k p_k J^k
\end{equation}
with coefficients $p_k\in\mathbb{C}$ of the polynomial $p$, as in the finite-dimensional setting a formal power series in $J$ is always finite. So any element of the produced algebra is represented as
\begin{equation}p(J)=p(\bigoplus_i J_{\lambda_i,m_i})=\bigoplus_i p(J_{\lambda_i,m_i})\end{equation} where the polynomials of Jordan blocks evaluate to
\begin{equation}p(J_{\lambda,m})=\begin{bmatrix}p(\lambda)&p^{(1)}(\lambda)&p^{(2)}(\lambda)/2!&\ddots&\ddots &\ddots&\ddots\\0&p(\lambda)&p^{(1)}(\lambda)&\ddots&\ddots& p^{(l)}(\lambda)/l!&\ddots\\0&0&p(\lambda)&\ddots&\ddots&\ddots&\ddots\\0&0&0&\ddots&\ddots&\ddots&\ddots\\0&0&0&0&p(\lambda)&p^{(1)}(\lambda)&p^{(2)}(\lambda)/2!\\0&0&0&0&0&p(\lambda)&p^{(1)}(\lambda)\\0&0&0&0&0&0&p(\lambda)\end{bmatrix}\end{equation}
with $p^{(l)}$ denoting the polynomial derivative of order $l$ in the subdiagonal of number $l$. The explicit form for any algebra element is therefore the diagonal block matrix composed of submatrices 
\begin{equation}
\resizebox{\textwidth}{!}{$\sum\limits_k\begin{bmatrix}p_k\lambda^k&p_kk\lambda^{k-1}&p_k(k^2-k)\lambda^{k-2}&\ddots&\ddots &\ddots&\ddots\\0&p_k\lambda^k&p_kk\lambda^{k-1}&\ddots&\ddots& p_kk!\lambda^{k-l}/(l!(k-l)!)&\ddots\\0&0&p_k\lambda^k&\ddots&\ddots&\ddots&\ddots\\0&0&0&\ddots&\ddots&\ddots&\ddots\\0&0&0&0&p_k\lambda^k&p_kk\lambda^{k-1}&p_k(k^2-k)\lambda^{k-2}\\0&0&0&0&0&p_k\lambda^k&p_kk\lambda^{k-1}\\0&0&0&0&0&0&p_k\lambda^k\end{bmatrix}$}
\end{equation}
that are of Toeplitz type and where the entries of the constant (sub)diagonals depend on the spectrum of the given generator $g$ as well as the specific choice of polynomial coefficients $p_k$ for the generated element.
\\\\
Possible are all polynomials with strictly positive powers $J^k$ ($k\in\mathbb{N}$ instead of $k\in\mathbb{N}_0$) and thus without the monomial $p_0 J^0$ due to our inability to assume the identity $I$ as a given generator. For nonzero spectra ($\lambda\neq 0$) the Jordan matrix is invertible and there exists a polynomial $i(J)=J^{-1}$ with $i(0)=0$. The matrix polynomial
\begin{equation}q_i(J):=\prod_{\lambda_i\neq \lambda_j} (J-\lambda_j Ji(J))^{m_j}\end{equation}
decomposes as the direct sum
\begin{equation} \bigoplus_j\prod_{\lambda_i\neq \lambda_j} (J_{\lambda_j,m_j}-\lambda_j J_{\lambda_j,m_j}i(J_{\lambda_j,m_j}))^{m_j}\end{equation}
where each of the summands $q_i(J_{\lambda_j,m_j})$ vanishes except for $q_i(J_{\lambda_i,m_i})$ as a product of invertible matrices (and thus itself invertible) whenever a $J$-cyclic subspace equals the full space. The respective polynomial $r_i(J_{\lambda_i,m_i}):=(q_i(J_{\lambda_i,m_i}))^{-1}$ for the inverse with $r_i(0)=0$ yields the matrix $r_i(J)q_i(J)$ that has the same block structure as $J$ with an identity submatrix $I_i$ in place of the Jordan block for eigenvalue $\lambda_i$ and all other blocks of the partition set to zero. 
\\\\
That way all identity operators $I_j$ can be obtained by linear combinations of strictly positive powers of $J$ via $r_j(J)q_j(J)$ for the generalized eigenspaces indexed by $j$. Identities for Jordan blocks with $\lambda = 0$ have to be ensured in a different way, for example via the square of respective Pauli matrices. All corresponding upper-triangular Toeplitz matrices $T_j$ are reached via the polynomial
\begin{equation}\
t(J) := \sum_j \sum_k (T_j)_{1k} (J-\lambda_jJi(J))^k r_j(J)q_j(J)
\end{equation}
where $(T_j)_{1k}$ refers to the first row of $T_j$ that fixes the entire upper-triangular Toeplitz submatrix block for the generalized eigenspace $j$. This proves that the direct sum of arbitrary upper-triangular Toeplitz blocks can really be generated via scalar multiplication as well as matrix addition and multiplication, since any choice for the $T_j$ is possible in the generating polynomial $t(J)$.\\\\
As transformations to the generalized eigenbasis of a generator constitute similiarity transformations, any algebra generated by operators is isomorphic to one generated in this special basis. This isomorphism enables an investigation of maximality for generated subalgebras in a basis-independent manner. Of particular advantage is the clearer picture of subspace invariance for generator sets with a matrix in Jordan canonical form, as this representation yields a suitable block structure that is preserved under the algebra operations and provides nilpotent matrix blocks $N_j:= J_{\lambda_j,m_j} - \lambda_j I_j$ (with nilpotency index $m_j$) useful for examining the patterns of invariant subspaces for  generator sets $S$: given any nonempty $S$-invariant subspace $V_J$, it is straightforward to verify via repeated action by those nilpotent blocks that $V_J$ must contain an element of the generalized eigenbasis, which in turn serves as the representation basis for all generators and therefore the element corresponds to a particular vertex $v$ of the associated digraph.
\\\\
To prove that full connectivity of that digraph implies $V_J=V$, it is enough to show that the rest of the generalized eigenbasis is also contained in $V_J$. Considering all vertices of the digraph that are the origin of an edge pointing to $v$, the respective nonzero entries of generator matrices corresponding to these edges can be manipulated via nilpotent blocks and linear combinations to confirm in recursive fashion the presence of all generalized eigenvectors in $V_J$. Hence the arbitrarily chosen $V_J$ contains a basis of the full space  and is thus the trivial subspace $V$, confirming the nonexistence of nontrivial invariant subspaces and the generation of the full operator algebra in case of strong digraph connectivity.
\subsection{Connectivity Webs}
It remains to clarify that the method is truly applicable to lattice systems of arbitrary size with $N$ spin sites and $D$ spatial dimensions, requiring the type of graph connectivity to stay unchanged. For that purpose we make use of visualizations in terms of "cobwebs" in order to investigate the threaded connectivity patterns that arise: the $N$ lattice sites induce a multi-spin Hilbert space of dimension $2^N$ and the respective digraph vertices are arranged with equal spacing on a circle in counter-clockwise order, such that an angle $(n-1)2\pi / 2^N$ is associated to vertex number $n\in\{1,...,2^N\}$.\\\\
Each addition of a new spin lattice site doubles the dimension of the state space and thus also the number of graph vertices, halving their angles in the circular arrangement. The vertex sets containing 
$2^N$ elements are depicted as concentric rings of circumference scaling with this cardinality, in an effort to make the steps $N \rightarrow N+1$ more transparent. The planar digraphs are drawn in blue, with arrowheads indicating the direction of the edges. All digraphs associated to the generator sets $\{H-i\sum L^*L/2, L\}$ are split into two parts: graph components due to the set $\{L\}$ of all Lindbladian jumps with the sum of all $L$ as adjacency matrix and components involving the Hamiltonian $H$ with adjacency matrix given by $H-i\sum L^*L/2$.\\\\
From the outset it is immediately perceptible in the connectivity webs that all graph components exhibit a disconnect between the first and last vertices; this is due to the fact that no circular boundary conditions (giving rise to the topology of $D$-dimensional tori) are imposed on the spin lattices and thus start and end regions of the graphs wrapped around circles are not directly woven together. Furthermore the graphical representation via cobwebs makes underlying symmetries of the connectivity patterns easily discernible: all graph components display rotational symmetry (modulo gaps due to lattice boundaries) corresponding to the translation invariance of the spin systems.
\\\\
The interplay of these two effects gives rise to characteristics for the evolution of the connectivity pattern with system size: the initial disjunctions for $N=2$ propagate outwards in the connectivity webs, doubling the number of disconnected direct neighbor vertices for each step $N$. This happens for both the Hamiltonian and Lindbladian components of our graph partition: the webs exhibit $2^{(N-1)}$  missing direct neighbor links for $\{L\}$ and $2^{(N-2)}$ for $H-i\sum L^*L/2$. At the same time a hierarchy of missing links for $k$-nearest neighbors emerges: once the number of sites is sufficient to allow for notions of $k$-locality, regularities appear on all scales. For example in the case of $k=2$ and $N > 2$, the number of missing links also doubles with each step and is equal to the number of $k=1$ disconnects. The patterns hold for all system sizes; TikZ pictures for two/three/four quantum lattice sites are included in this documentation.
\\\\
In addition it is notable that all digraph components for $\{L\}$ are unidirectional while all digraph components for $H-i\sum L^*L/2$ are bidirectional. Figures 1-3 show the directed graphs all possessing a clockwise orientation with no bidirectional links; Figures 4-6 show the entirely reversible character of the graphs with no unidirectional links: vertices are always connected in both ways or not at all. Figures 1-6 also reveal that vertices on opposite sides of the circles are always directly connected, again uni/bi-directional for Lindbladian/Hamiltonian components respectively.
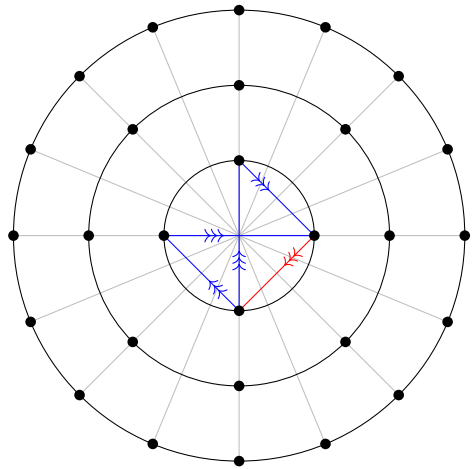
\begin{figure}[h!]
\centering
\begin{tikzpicture}[decoration={markings, mark=at position 0.4 with {\arrow{>>>}}}]

  \def\outerRadius{3}
  \def\middleRadius{2}
  \def\innerRadius{1}

  \def\innerPoints{4}
  \def\middlePoints{8}
  \def\outerPoints{16}

  \foreach \angle in {0,22.5,...,337.5} {
    \draw[lightgray, thin] (0,0) -- (\angle:\outerRadius);
  }  
 
  
  \foreach \i/\j/\val in {1/1/0, 1/2/0, 1/3/0, 1/4/0,
                           2/1/1, 2/2/0, 2/3/0, 2/4/0,
                           3/1/1, 3/2/0, 3/3/0, 3/4/0,
                           4/1/0, 4/2/1, 4/3/1, 4/4/0} {
    \ifnum\val>0
      \draw[postaction={decorate}, blue, thin] ({(\i * 90-90)}:{\innerRadius}) -- ({(\j * 90-90)}:{\innerRadius});
    \fi
  }
  \draw[postaction={decorate}, red, thin] ({0}:{\innerRadius}) -- ({270}:{\innerRadius});
  \draw (0,0) circle (\outerRadius);
  \draw (0,0) circle (\middleRadius);
  \draw (0,0) circle (\innerRadius);

  \foreach \angle in {0,90,180,270} {
    \fill (\angle:\innerRadius) circle (2pt);
  }

  \foreach \angle in {0,45,...,315} {
    \fill (\angle:\middleRadius) circle (2pt);
  }

  \foreach \angle in {0,22.5,...,337.5} {
    \fill (\angle:\outerRadius) circle (2pt);
  }
 \end{tikzpicture}
\caption{Web of Connectivity for $\{L\}$ with 2 Sites}
\end{figure}
\begin{figure}[h!]
\centering
\begin{tikzpicture}[decoration={markings, mark=at position 0.4 with {\arrow{>>>}}}]

  \def\outerRadius{3}
  \def\middleRadius{2}
  \def\innerRadius{1}

  \def\innerPoints{4}
  \def\middlePoints{8}
  \def\outerPoints{16}

  \foreach \angle in {0,22.5,...,337.5} {
    \draw[lightgray, thin] (0,0) -- (\angle:\outerRadius);
  }

 \foreach \i/\j/\val in {1/1/0, 1/2/0, 1/3/0, 1/4/0, 1/5/0, 1/6/0, 1/7/0,
                        2/1/1, 2/2/0, 2/3/0, 2/4/0, 2/5/0, 2/6/0, 2/7/0,
                        3/1/1, 3/2/0, 3/3/0, 3/4/0, 3/5/0, 3/6/0, 3/7/0,
                        4/1/0, 4/2/1, 4/3/1, 4/4/0, 4/5/0, 4/6/0, 4/7/0,
                        5/1/1, 5/2/0, 5/3/0, 5/4/0, 5/5/0, 5/6/0, 5/7/0,
                        6/1/0, 6/2/1, 6/3/0, 6/4/0, 6/5/1, 6/6/0, 6/7/0,
                        7/1/0, 7/2/0, 7/3/1, 7/4/0, 7/5/1, 7/6/0, 7/7/0,
                        8/1/0, 8/2/0, 8/3/0, 8/4/1, 8/5/0, 8/6/1, 8/7/1,8/8/0}
{
    \ifnum\val>0
      \draw[postaction={decorate}, blue, thin] ({(\i * 45-45)}:{\middleRadius}) -- ({(\j * 45-45)}:{\middleRadius});
    \fi
  }
   \draw[postaction={decorate}, red, thin] ({0}:{\middleRadius}) -- ({315}:{\middleRadius});
  \draw (0,0) circle (\outerRadius);
  \draw (0,0) circle (\middleRadius);
  \draw (0,0) circle (\innerRadius);

  \foreach \angle in {0,90,180,270} {
    \fill (\angle:\innerRadius) circle (2pt);
  }

  \foreach \angle in {0,45,...,315} {
    \fill (\angle:\middleRadius) circle (2pt);
  }

  \foreach \angle in {0,22.5,...,337.5} {
    \fill (\angle:\outerRadius) circle (2pt);
  }
\end{tikzpicture}
\caption{Web of Connectivity for $\{L\}$ with 3 Sites}
\end{figure}
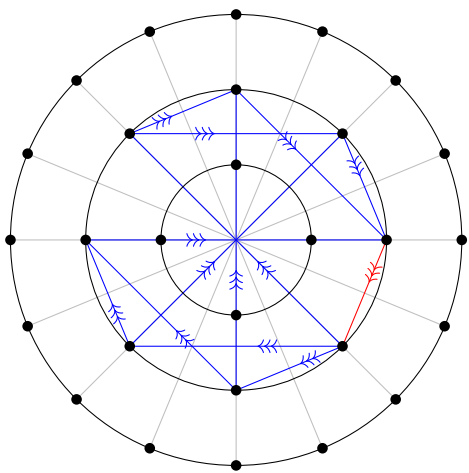
\begin{figure}[h!]
\centering
\begin{tikzpicture}[decoration={markings, mark=at position 0.4 with {\arrow{>>>}}}]

  \def\outerRadius{3}
  \def\middleRadius{2}
  \def\innerRadius{1}

  \def\innerPoints{4}
  \def\middlePoints{8}
  \def\outerPoints{16}

  \foreach \angle in {0,22.5,...,337.5} {
    \draw[lightgray, thin] (0,0) -- (\angle:\outerRadius);
  }

\foreach \i/\j/\val in {1/1/0, 1/2/0, 1/3/0, 1/4/0, 1/5/0, 1/6/0, 1/7/0, 1/8/0, 1/9/0, 1/10/0, 1/11/0, 1/12/0, 1/13/0, 1/14/0, 1/15/0, 1/16/0,
                         2/1/1, 2/2/0, 2/3/0, 2/4/0, 2/5/0, 2/6/0, 2/7/0, 2/8/0, 2/9/0, 2/10/0, 2/11/0, 2/12/0, 2/13/0, 2/14/0, 2/15/0, 2/16/0,
                         3/1/1, 3/2/0, 3/3/0, 3/4/0, 3/5/0, 3/6/0, 3/7/0, 3/8/0, 3/9/0, 3/10/0, 3/11/0, 3/12/0, 3/13/0, 3/14/0, 3/15/0, 3/16/0,
                         4/1/0, 4/2/1, 4/3/1, 4/4/0, 4/5/0, 4/6/0, 4/7/0, 4/8/0, 4/9/0, 4/10/0, 4/11/0, 4/12/0, 4/13/0, 4/14/0, 4/15/0, 4/16/0,
                         5/1/1, 5/2/0, 5/3/0, 5/4/0, 5/5/0, 5/6/0, 5/7/0, 5/8/0, 5/9/0, 5/10/0, 5/11/0, 5/12/0, 5/13/0, 5/14/0, 5/15/0, 5/16/0,
                         6/1/0, 6/2/1, 6/3/0, 6/4/0, 6/5/1, 6/6/0, 6/7/0, 6/8/0, 6/9/0, 6/10/0, 6/11/0, 6/12/0, 6/13/0, 6/14/0, 6/15/0, 6/16/0,
                         7/1/0, 7/2/0, 7/3/1, 7/4/0, 7/5/1, 7/6/0, 7/7/0, 7/8/0, 7/9/0, 7/10/0, 7/11/0, 7/12/0, 7/13/0, 7/14/0, 7/15/0, 7/16/0,
                         8/1/0, 8/2/0, 8/3/0, 8/4/1, 8/5/0, 8/6/1, 8/7/1, 8/8/0, 8/9/0, 8/10/0, 8/11/0, 8/12/0, 8/13/0, 8/14/0, 8/15/0, 8/16/0,
                         9/1/1, 9/2/0, 9/3/0, 9/4/0, 9/5/0, 9/6/0, 9/7/0, 9/8/0, 9/9/0, 9/10/0, 9/11/0, 9/12/0, 9/13/0, 9/14/0, 9/15/0, 9/16/0,
                         10/1/0, 10/2/1, 10/3/0, 10/4/0, 10/5/0, 10/6/0, 10/7/0, 10/8/0, 10/9/1, 10/10/0, 10/11/0, 10/12/0, 10/13/0, 10/14/0, 10/15/0, 10/16/0,
                         11/1/0, 11/2/0, 11/3/1, 11/4/0, 11/5/0, 11/6/0, 11/7/0, 11/8/0, 11/9/1, 11/10/0, 11/11/0, 11/12/0, 11/13/0, 11/14/0, 11/15/0, 11/16/0,
                         12/1/0, 12/2/0, 12/3/0, 12/4/1, 12/5/0, 12/6/0, 12/7/0, 12/8/0, 12/9/0, 12/10/1, 12/11/1, 12/12/0, 12/13/0, 12/14/0, 12/15/0, 12/16/0,
                         13/1/0, 13/2/0, 13/3/0, 13/4/0, 13/5/1, 13/6/0, 13/7/0, 13/8/0, 13/9/1, 13/10/0, 13/11/0, 13/12/0, 13/13/0, 13/14/0, 13/15/0, 13/16/0,
                         14/1/0, 14/2/0, 14/3/0, 14/4/0, 14/5/0, 14/6/1, 14/7/0, 14/8/0, 14/9/0, 14/10/1, 14/11/0, 14/12/0, 14/13/1, 14/14/0, 14/15/0, 14/16/0,
                         15/1/0, 15/2/0, 15/3/0, 15/4/0, 15/5/0, 15/6/0, 15/7/1, 15/8/0, 15/9/0, 15/10/0, 15/11/1, 15/12/0, 15/13/1, 15/14/0, 15/15/0, 15/16/0,
                         16/1/0, 16/2/0, 16/3/0, 16/4/0, 16/5/0, 16/6/0, 16/7/0, 16/8/1, 16/9/0, 16/10/0, 16/11/0, 16/12/1, 16/13/0, 16/14/1, 16/15/1, 16/16/0} {
  \ifnum\val>0
    \draw[postaction={decorate}, blue, thin] ({(\i * 22.5-22.5)}:{\outerRadius}) -- ({(\j * 22.5-22.5)}:{\outerRadius});
  \fi
}
  \draw[postaction={decorate}, red, thin] ({0}:{\outerRadius}) -- ({338.5}:{\outerRadius});
  \draw (0,0) circle (\outerRadius);
  \draw (0,0) circle (\middleRadius);
  \draw (0,0) circle (\innerRadius);

  \foreach \angle in {0,90,180,270} {
    \fill (\angle:\innerRadius) circle (2pt);
  }

  \foreach \angle in {0,45,...,315} {
    \fill (\angle:\middleRadius) circle (2pt);
  }

  \foreach \angle in {0,22.5,...,337.5} {
    \fill (\angle:\outerRadius) circle (2pt);
  }\end{tikzpicture}
\caption{Web of Connectivity for $\{L\}$ with 4 Sites}
\end{figure}
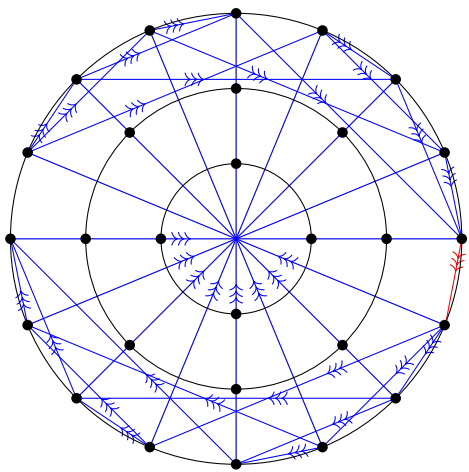
\begin{figure}[h!]
\centering
\begin{tikzpicture}[decoration={markings, mark=at position 0.4 with {\arrow{>>>}}}]

  \def\outerRadius{3}
  \def\middleRadius{2}
  \def\innerRadius{1}

  \def\innerPoints{4}
  \def\middlePoints{8}
  \def\outerPoints{16}

  \foreach \angle in {0,22.5,...,337.5} {
    \draw[lightgray, thin] (0,0) -- (\angle:\outerRadius);
  }  
 
  
  \foreach \i/\j/\val in {1/1/1, 1/2/1, 1/3/1, 1/4/0,
                           2/1/1, 2/2/1, 2/3/1, 2/4/1,
                           3/1/1, 3/2/1, 3/3/1, 3/4/1,
                           4/1/0, 4/2/1, 4/3/1, 4/4/1} {
    \ifnum\val>0
      \draw[postaction={decorate}, blue, thin] ({(\i * 90-90)}:{\innerRadius}) -- ({(\j * 90-90)}:{\innerRadius});
    \fi
  }
  \draw (0,0) circle (\outerRadius);
  \draw (0,0) circle (\middleRadius);
  \draw (0,0) circle (\innerRadius);

  \foreach \angle in {0,90,180,270} {
    \fill (\angle:\innerRadius) circle (2pt);
  }

  \foreach \angle in {0,45,...,315} {
    \fill (\angle:\middleRadius) circle (2pt);
  }

  \foreach \angle in {0,22.5,...,337.5} {
    \fill (\angle:\outerRadius) circle (2pt);
  }
 \end{tikzpicture}
\caption{Web of Connectivity for $H-i\sum L^*L/2$ with 2 Sites}
\end{figure}
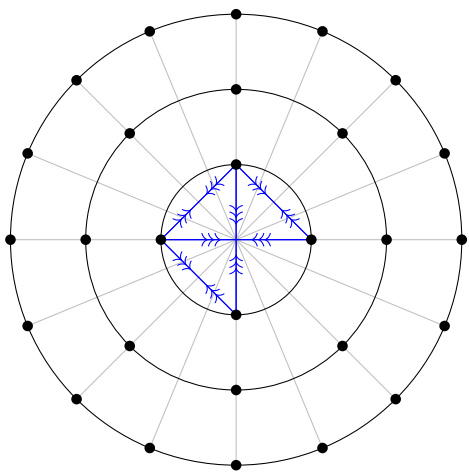
\begin{figure}[h!]
\centering
\begin{tikzpicture}[decoration={markings, mark=at position 0.4 with {\arrow{>>>}}}]

  \def\outerRadius{3}
  \def\middleRadius{2}
  \def\innerRadius{1}

  \def\innerPoints{4}
  \def\middlePoints{8}
  \def\outerPoints{16}

  \foreach \angle in {0,22.5,...,337.5} {
    \draw[lightgray, thin] (0,0) -- (\angle:\outerRadius);
  }

 \foreach \i/\j/\val in {1/1/1, 1/2/1, 1/3/1, 1/4/0, 1/5/1, 1/6/0, 1/7/0, 1/8/0,
                        2/1/1, 2/2/1, 2/3/1, 2/4/1, 2/5/1, 2/6/1, 2/7/0, 2/8/0,
                        3/1/1, 3/2/1, 3/3/1, 3/4/1, 3/5/1, 3/6/0, 3/7/1, 3/8/0,
                        4/1/0, 4/2/1, 4/3/1, 4/4/1, 4/5/0, 4/6/1, 4/7/1, 4/8/1,
                        5/1/1, 5/2/1, 5/3/1, 5/4/0, 5/5/1, 5/6/1, 5/7/1, 5/8/0,
                        6/1/0, 6/2/1, 6/3/0, 6/4/1, 6/5/1, 6/6/1, 6/7/1, 6/8/1,
                        7/1/0, 7/2/0, 7/3/1, 7/4/1, 7/5/1, 7/6/1, 7/7/1,7/8/1,
                        8/1/0, 8/2/0, 8/3/0, 8/4/1, 8/5/0, 8/6/1, 8/7/1,8/8/1}
{
    \ifnum\val>0
      \draw[postaction={decorate}, blue, thin] ({(\i * 45-45)}:{\middleRadius}) -- ({(\j * 45-45)}:{\middleRadius});
    \fi
  }
  \draw (0,0) circle (\outerRadius);
  \draw (0,0) circle (\middleRadius);
  \draw (0,0) circle (\innerRadius);

  \foreach \angle in {0,90,180,270} {
    \fill (\angle:\innerRadius) circle (2pt);
  }

  \foreach \angle in {0,45,...,315} {
    \fill (\angle:\middleRadius) circle (2pt);
  }

  \foreach \angle in {0,22.5,...,337.5} {
    \fill (\angle:\outerRadius) circle (2pt);
  }
\end{tikzpicture}
\caption{Web of Connectivity for $H-i\sum L^*L/2$ with 3 Sites}
\end{figure}
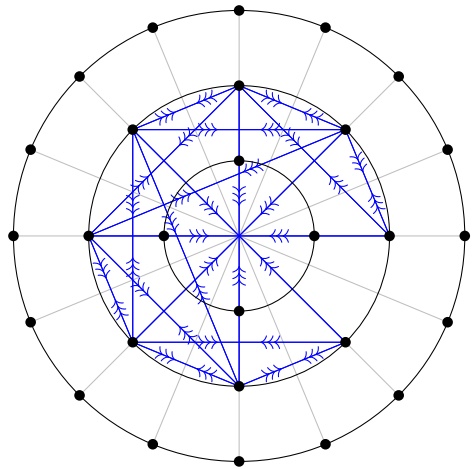
\begin{figure}[h!]
\centering
\begin{tikzpicture}[decoration={markings, mark=at position 0.45 with {\arrow{>>>}}}, scale=1.5]
  \def\outerRadius{3}
  \def\middleRadius{2}
  \def\innerRadius{1}

  \def\innerPoints{4}
  \def\middlePoints{8}
  \def\outerPoints{16}

  \foreach \angle in {0,22.5,...,337.5} {
    \draw[lightgray, ultra thin] (0,0) -- (\angle:\outerRadius);
  }

\foreach \i/\j/\val in {1/1/1, 1/2/1, 1/3/1, 1/4/0, 1/5/1, 1/6/0, 1/7/0, 1/8/0, 1/9/1, 1/10/0, 1/11/0, 1/12/0, 1/13/0, 1/14/0, 1/15/0, 1/16/0,
                         2/1/1, 2/2/1, 2/3/1, 2/4/1, 2/5/0, 2/6/1, 2/7/0, 2/8/0, 2/9/1, 2/10/1, 2/11/0, 2/12/0, 2/13/0, 2/14/0, 2/15/0, 2/16/0,
                         3/1/1, 3/2/1, 3/3/1, 3/4/1, 3/5/1, 3/6/0, 3/7/1, 3/8/0, 3/9/0, 3/10/0, 3/11/1, 3/12/0, 3/13/0, 3/14/0, 3/15/0, 3/16/0,
                         4/1/0, 4/2/1, 4/3/1, 4/4/1, 4/5/0, 4/6/1, 4/7/0, 4/8/1, 4/9/0, 4/10/0, 4/11/1, 4/12/1, 4/13/0, 4/14/0, 4/15/0, 4/16/0,
                         5/1/1, 5/2/0, 5/3/1, 5/4/0, 5/5/1, 5/6/1, 5/7/1, 5/8/0, 5/9/1, 5/10/0, 5/11/0, 5/12/0, 5/13/1, 5/14/0, 5/15/0, 5/16/0,
                         6/1/0, 6/2/1, 6/3/0, 6/4/1, 6/5/1, 6/6/1, 6/7/1, 6/8/1, 6/9/0, 6/10/1, 6/11/0, 6/12/0, 6/13/1, 6/14/1, 6/15/0, 6/16/0,
                         7/1/0, 7/2/0, 7/3/1, 7/4/0, 7/5/1, 7/6/1, 7/7/1, 7/8/1, 7/9/0, 7/10/0, 7/11/1, 7/12/0, 7/13/0, 7/14/0, 7/15/1, 7/16/0,
                         8/1/0, 8/2/0, 8/3/0, 8/4/1, 8/5/0, 8/6/1, 8/7/1, 8/8/1, 8/9/0, 8/10/0, 8/11/0, 8/12/1, 8/13/0, 8/14/0, 8/15/1, 8/16/1,
                         9/1/1, 9/2/1, 9/3/0, 9/4/0, 9/5/1, 9/6/0, 9/7/0, 9/8/0, 9/9/1, 9/10/1, 9/11/1, 9/12/0, 9/13/1, 9/14/0, 9/15/0, 9/16/0,
                         10/1/0, 10/2/1, 10/3/0, 10/4/0, 10/5/0, 10/6/1, 10/7/0, 10/8/0, 10/9/1, 10/10/1, 10/11/1, 10/12/1, 10/13/0, 10/14/1, 10/15/0, 10/16/0,
                         11/1/0, 11/2/0, 11/3/1, 11/4/1, 11/5/0, 11/6/0, 11/7/1, 11/8/0, 11/9/1, 11/10/1, 11/11/1, 11/12/1, 11/13/1, 11/14/0, 11/15/1, 11/16/0,
                         12/1/0, 12/2/0, 12/3/0, 12/4/1, 12/5/0, 12/6/0, 12/7/0, 12/8/1, 12/9/0, 12/10/1, 12/11/1, 12/12/1, 12/13/0, 12/14/1, 12/15/0, 12/16/1,
                         13/1/0, 13/2/0, 13/3/0, 13/4/0, 13/5/1, 13/6/1, 13/7/0, 13/8/0, 13/9/1, 13/10/0, 13/11/1, 13/12/0, 13/13/1, 13/14/1, 13/15/1, 13/16/0,
                         14/1/0, 14/2/0, 14/3/0, 14/4/0, 14/5/0, 14/6/1, 14/7/0, 14/8/0, 14/9/0, 14/10/1, 14/11/0, 14/12/1, 14/13/1, 14/14/1, 14/15/1, 14/16/1,
                         15/1/0, 15/2/0, 15/3/0, 15/4/0, 15/5/0, 15/6/0, 15/7/1, 15/8/1, 15/9/0, 15/10/0, 15/11/1, 15/12/0, 15/13/1, 15/14/1, 15/15/1, 15/16/1,
                         16/1/0, 16/2/0, 16/3/0, 16/4/0, 16/5/0, 16/6/0, 16/7/0, 16/8/1, 16/9/0, 16/10/0, 16/11/0, 16/12/1, 16/13/0, 16/14/1, 16/15/1, 16/16/1} {
  \ifnum\val>0
    \draw[postaction={decorate}, blue, thin] ({(\i * 22.5-22.5)}:{\outerRadius}) -- ({(\j * 22.5-22.5)}:{\outerRadius});
  \fi
}
  \draw (0,0) circle (\outerRadius);
  \draw (0,0) circle (\middleRadius);
  \draw (0,0) circle (\innerRadius);

  \foreach \angle in {0,90,180,270} {
    \fill (\angle:\innerRadius) circle (2pt);
  }

  \foreach \angle in {0,45,...,315} {
    \fill (\angle:\middleRadius) circle (2pt);
  }

  \foreach \angle in {0,22.5,...,337.5} {
    \fill (\angle:\outerRadius) circle (2pt);
  }\end{tikzpicture}
\caption{Web of Connectivity for $H-i\sum L^*L/2$ with 4 Sites}
\end{figure}
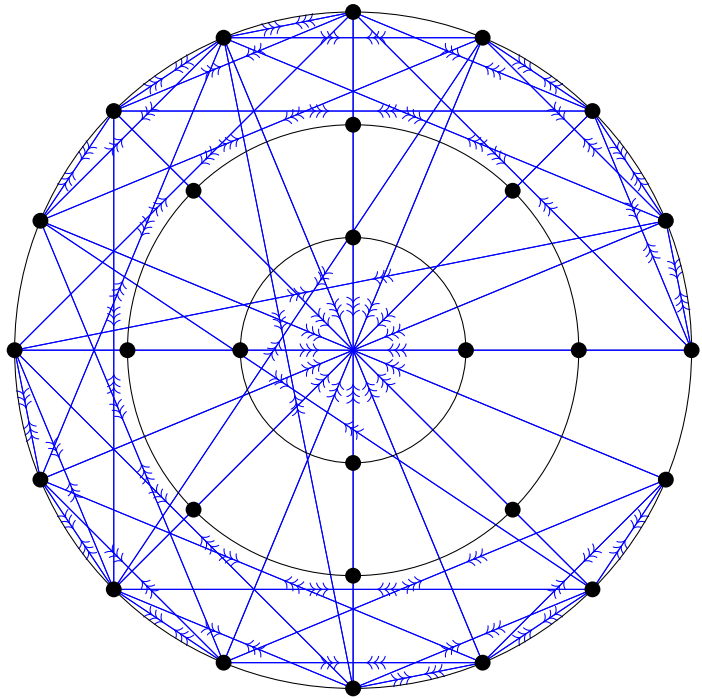
\subsection{Graph Self-Similarity}
The interwoven connectivity webs will now be scrutinized further in an algebraic way to explain their scale-invariance or self-similarity: the same operations involved in each step of growing system size $N$, namely matrix Kronecker multiplication and matrix addition, translate to different kinds of repeated digraph concatenations. In an effort to disentangle those, it is best to concentrate first on the components for $\{L\}$ due to the lack of negative interference in comparison to $H-i\sum L^*L/2$.
\\\\
Strong connectivity is guaranteed whenever a closed walk visiting all vertices can be found within the graph. Such cycles are possible for all sizes with the mere addition of only one single directed edge from the first to the last vertex, marked with a red arrow in Figures 1-3. This means that the Lindbladian jump operators give rise to strong connectivity almost on their own, irrespective of the specific Hamiltonian except for the condition that $H-i\sum L^*L/2$ has to contribute this singular missing edge. That way strong connectivity is ensured via generators $H-i\sum L^*L/2$ that are strongly connected themselves and/or provide this one directed edge from vertex number $1$ to vertex number $2^N$.
\\\\
The ratios between broken/missing links and preserved connections in the $k$-hierarchy are crucial for the quality of strong connectivity to persevere and prevent the breakup of the graph into individual strongly connected pieces. The proliferation of newly woven links has to overcompensate the loss of connections holding the cobweb together; an examination of this evolution is possible via the adjacency matrix encoding all graph edges for $\{L\}$ defined as
\begin{equation} A_N := \sum_{i=1}^N \sigma_i^{-} = \sum_{i=1}^N I^{\otimes (i-1)}\otimes \sigma^- \otimes I^{\otimes(N-i)}
\end{equation}
for $N$ lattice sites. Calculating the sum term by term, it becomes clear that this summation gives rise to a self-similar sequence in $N$: the first summand $\sigma_1^-=\sigma^- \otimes I^{\otimes(N-1)}$ yields a block matrix partitioned into four matrix blocks of dimension $2^N/2$ with an identity on the lower left as the only nonzero block; this partitioning is continued via the second summand $\sigma_2^-=I\otimes \sigma^- \otimes I^{\otimes(N-2)}$ dividing the matrix into $2^4$ blocks of dimension $2^N/4$ and all subsequent summands $\sigma_k^{-} = I^{\otimes (k-1)}\otimes \sigma^- \otimes I^{\otimes(N-k)}$, each contributing $2^{2k}$ matrix blocks of dimension $2^N/{(2k)}$ with $2k$ identity blocks and $2^{2k}-2k$ zero blocks.
\\\\
As a consequence of this summation pattern, the full adjacency matrix $A_N$ exhibits the self-similar shape of a fractal with perfect scale-invariance in the thermodynamic limit $N \rightarrow \infty$. Every step $A_N \rightarrow A_{N+1}$ corresponds to placing two copies of $ A_N$ as the diagonal blocks of $A_{N+1}$ and an identity block into the lower-left corner, shown in Figures 7 and 8.
\begin{figure}[h!]
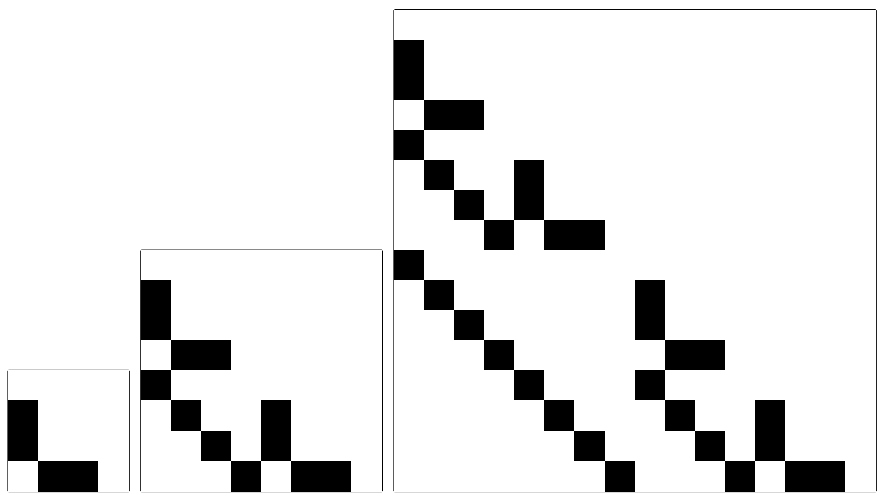

\centering

\caption{Adjacency Matrices for $\{L\}$ with 2 to 4 Sites}
\end{figure}
\begin{figure}[h!]
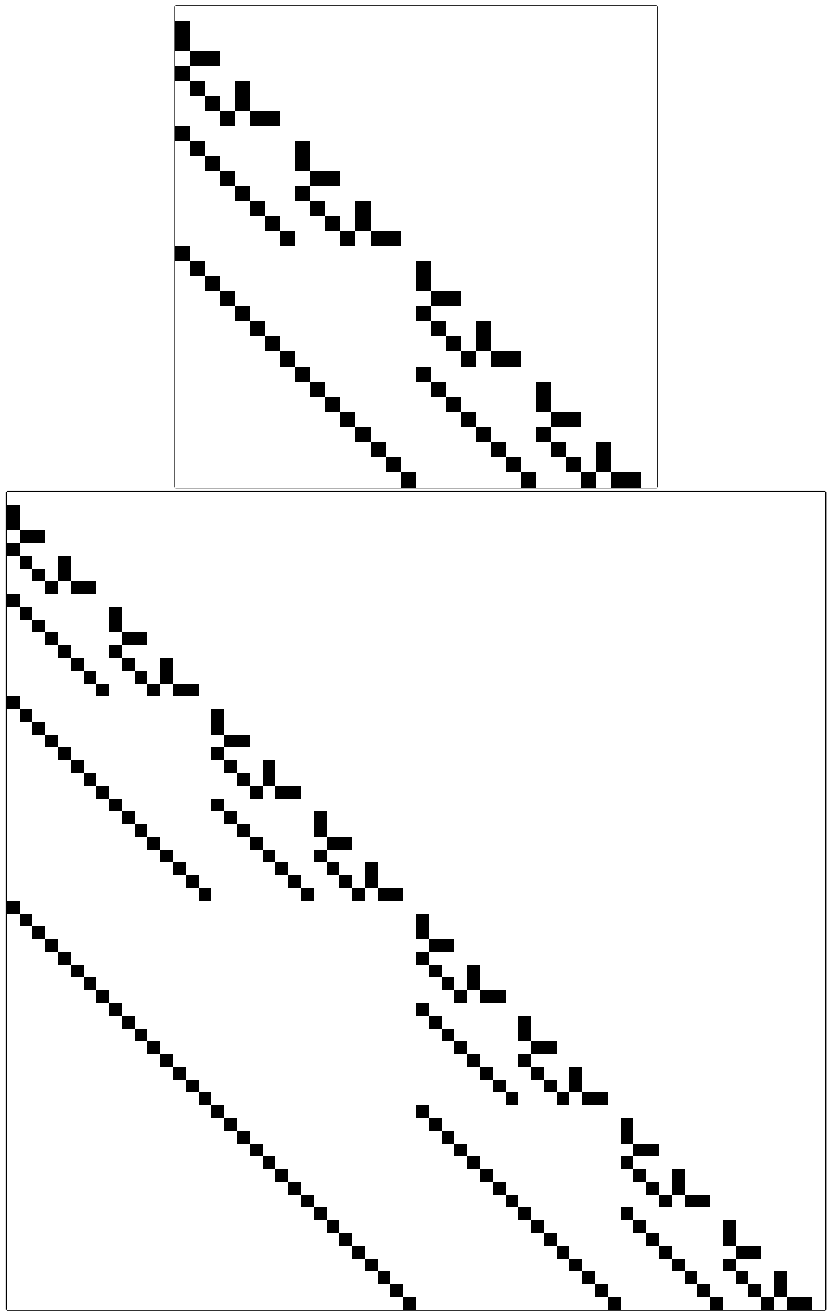

\centering
%
\caption{Adjacency Matrices for $\{L\}$ with 5 and 6 Sites}
\end{figure}
The fractal nature of the adjacency matrix $A_{N\rightarrow \infty}$ carries over to the digraph and explains all observed features: the basic geometrical shape for $N=2$ of two triangles at opposing sides of the connectivity web is repeated over and over again for higher numbers of lattice sites and connections are woven in correspondence with the placement of identity (one copy) and adjacency matrices (two copies) of the previous step. The unidirectional orientation of the graph is due to the zero upper triangle of $A_N$ for all $N$. The diagonal of an adjacency matrix refers to self-loops of graph vertices and thus plays no role in connectivity studies; it was therefore not included in the definition of $A_N$.
\\\\
From this construction it is immediately clear that no established connections are ever lost in the steps $N \rightarrow N+1$ due to the nested procedure with complete copies of the prior adjacency matrix. The conservation of the previous skeleton is exemplified in the visualization of Figure 9.
\begin{figure}[h!]
\centering
\begin{tikzpicture}[scale=1.5]
\tikzstyle{vertex} = [circle, draw, fill=black, inner sep=0pt, minimum size=0.1cm]
\def\outerRadius{2.5}
\draw[lightgray, thin] (0,0) circle (\outerRadius);
\node[vertex] (2_0) at (2.5,0.0) {};
\node[vertex] (2_1) at (1.5308084989341916e-16,2.5) {};
\node[vertex] (2_2) at (-2.5,3.061616997868383e-16) {};
\node[vertex] (2_3) at (-4.592425496802574e-16,-2.5) {};
\draw[line width=0.9000000000000001mm, color=blue] (2_1) -- (2_0);
\draw[line width=0.9000000000000001mm, color=blue] (2_2) -- (2_0);
\draw[line width=0.9000000000000001mm, color=blue] (2_3) -- (2_1);
\draw[line width=0.9000000000000001mm, color=blue] (2_3) -- (2_2);
\node[vertex] (3_0) at (2.5,0.0) {};
\node[vertex] (3_1) at (1.7677669529663689,1.7677669529663687) {};
\node[vertex] (3_2) at (1.5308084989341916e-16,2.5) {};
\node[vertex] (3_3) at (-1.7677669529663687,1.7677669529663689) {};
\node[vertex] (3_4) at (-2.5,3.061616997868383e-16) {};
\node[vertex] (3_5) at (-1.7677669529663693,-1.7677669529663687) {};
\node[vertex] (3_6) at (-4.592425496802574e-16,-2.5) {};
\node[vertex] (3_7) at (1.7677669529663684,-1.7677669529663693) {};
\draw[line width=0.6000000000000001mm, color=green] (3_1) -- (3_0);
\draw[line width=0.6000000000000001mm, color=green] (3_2) -- (3_0);
\draw[line width=0.6000000000000001mm, color=green] (3_3) -- (3_1);
\draw[line width=0.6000000000000001mm, color=green] (3_3) -- (3_2);
\draw[line width=0.6000000000000001mm, color=green] (3_4) -- (3_0);
\draw[line width=0.6000000000000001mm, color=green] (3_5) -- (3_1);
\draw[line width=0.6000000000000001mm, color=green] (3_5) -- (3_4);
\draw[line width=0.6000000000000001mm, color=green] (3_6) -- (3_2);
\draw[line width=0.6000000000000001mm, color=green] (3_6) -- (3_4);
\draw[line width=0.6000000000000001mm, color=green] (3_7) -- (3_3);
\draw[line width=0.6000000000000001mm, color=green] (3_7) -- (3_5);
\draw[line width=0.6000000000000001mm, color=green] (3_7) -- (3_6);
\node[vertex] (4_0) at (2.5,0.0) {};
\node[vertex] (4_1) at (2.309698831278217,0.9567085809127245) {};
\node[vertex] (4_2) at (1.7677669529663689,1.7677669529663687) {};
\node[vertex] (4_3) at (0.9567085809127246,2.309698831278217) {};
\node[vertex] (4_4) at (1.5308084989341916e-16,2.5) {};
\node[vertex] (4_5) at (-0.9567085809127243,2.309698831278217) {};
\node[vertex] (4_6) at (-1.7677669529663687,1.7677669529663689) {};
\node[vertex] (4_7) at (-2.309698831278217,0.9567085809127247) {};
\node[vertex] (4_8) at (-2.5,3.061616997868383e-16) {};
\node[vertex] (4_9) at (-2.309698831278217,-0.9567085809127241) {};
\node[vertex] (4_10) at (-1.7677669529663693,-1.7677669529663687) {};
\node[vertex] (4_11) at (-0.9567085809127238,-2.309698831278217) {};
\node[vertex] (4_12) at (-4.592425496802574e-16,-2.5) {};
\node[vertex] (4_13) at (0.956708580912725,-2.3096988312782165) {};
\node[vertex] (4_14) at (1.7677669529663684,-1.7677669529663693) {};
\node[vertex] (4_15) at (2.309698831278217,-0.9567085809127239) {};
\draw[line width=0.2mm, color=red] (4_1) -- (4_0);
\draw[line width=0.2mm, color=red] (4_2) -- (4_0);
\draw[line width=0.2mm, color=red] (4_3) -- (4_1);
\draw[line width=0.2mm, color=red] (4_3) -- (4_2);
\draw[line width=0.2mm, color=red] (4_4) -- (4_0);
\draw[line width=0.2mm, color=red] (4_5) -- (4_1);
\draw[line width=0.2mm, color=red] (4_5) -- (4_4);
\draw[line width=0.2mm, color=red] (4_6) -- (4_2);
\draw[line width=0.2mm, color=red] (4_6) -- (4_4);
\draw[line width=0.2mm, color=red] (4_7) -- (4_3);
\draw[line width=0.2mm, color=red] (4_7) -- (4_5);
\draw[line width=0.2mm, color=red] (4_7) -- (4_6);
\draw[line width=0.2mm, color=red] (4_8) -- (4_0);
\draw[line width=0.2mm, color=red] (4_9) -- (4_1);
\draw[line width=0.2mm, color=red] (4_9) -- (4_8);
\draw[line width=0.2mm, color=red] (4_10) -- (4_2);
\draw[line width=0.2mm, color=red] (4_10) -- (4_8);
\draw[line width=0.2mm, color=red] (4_11) -- (4_3);
\draw[line width=0.2mm, color=red] (4_11) -- (4_9);
\draw[line width=0.2mm, color=red] (4_11) -- (4_10);
\draw[line width=0.2mm, color=red] (4_12) -- (4_4);
\draw[line width=0.2mm, color=red] (4_12) -- (4_8);
\draw[line width=0.2mm, color=red] (4_13) -- (4_5);
\draw[line width=0.2mm, color=red] (4_13) -- (4_9);
\draw[line width=0.2mm, color=red] (4_13) -- (4_12);
\draw[line width=0.2mm, color=red] (4_14) -- (4_6);
\draw[line width=0.2mm, color=red] (4_14) -- (4_10);
\draw[line width=0.2mm, color=red] (4_14) -- (4_12);
\draw[line width=0.2mm, color=red] (4_15) -- (4_7);
\draw[line width=0.2mm, color=red] (4_15) -- (4_11);
\draw[line width=0.2mm, color=red] (4_15) -- (4_13);
\draw[line width=0.2mm, color=red] (4_15) -- (4_14);
\end{tikzpicture}
\caption{Connectivity for 2/3/4 Sites (blue/green/red)}
\end{figure}
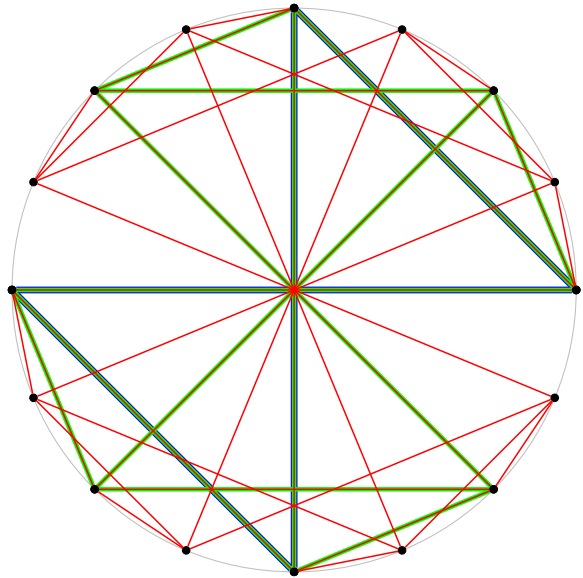
\\\\
The equivalence between strong connectivity and the existence of a closed path comprising the entire vertex set implies that the above claim about a single missing link (arrow from first to last vertex) for all $N$ causing every digraph to be strongly connected can be proven by giving an explicit recipe for such a path visiting all vertices; this recipe will be presented below, but there is an even more elegant way to show strong connectivity:
\\\\
The number of all paths with $p$ edges is encoded in the powers $A_N^p$ of the respective adjacency matrices: the matrix entry $(A_N^p)_{kl}$ gives the number of paths from vertex $k$ to vertex $l$ traversing $p-1$ vertices, excluding $k$ and $l$. For example, $A^2_2= (\sigma^-\otimes I + I\otimes \sigma^-)^2=2(\sigma^- \otimes \sigma^-)$ and all higher powers of $A_2$ vanish due to the nilpotency of $\sigma^- $. This means there are exactly two paths with two edges from the last to the first vertex, but no other paths: neither with two edges between other vertices, nor with more edges between any vertices. Similarly, $A^3_3= 6(\sigma^- \otimes \sigma^- \otimes \sigma^-)$ and all higher powers of $A_3$ vanish; thus there exist six paths with three edges between the last and first vertex. 
\\\\
Other paths with the same or higher number of edges fail due to the lack of a further connection. The sparsity of the adjacency matrices (defined as the percentage of non-zero elements) could be suspected as the reason for a lack of strong connectivity, but in fact the true culprit is their nilpotent character: due to the nilpotency of all $A_N$ as strictly lower-triangular matrices, no path visiting all vertices can exist. In fact, an even stronger statement is true: the digraph for $A_N$ contains no path with more than $N$ edges, much less than the needed minimum for connectivity.
\\\\
This situation changes drastically by "nilpotency breaking" in the empty upper triangle of $A_N$ via the addition of a nonzero matrix element in the first row and last column. With that modification higher powers no longer vanish; the deformed matrix will subsequently be called $C_N:=A_N+B_N$ with the deformation $B_N:=\delta_{1,2^N}$ given by a Kronecker delta $\delta$. As the powers of adjacency matrices represent the number of paths with a specific length between pairs of vertices, a sufficient (and necessary) criterion for strong connectivity is the existence of a matrix with only nonzero entries obtained as the sum 
\begin{equation}
\sum_k C^k_N
\end{equation}
In that (and only that) case is a directed path ensured for any ordered vertex pair of the altered digraph. As paths of length $2^N$ (or longer) necessarily include cycles, only powers up to $2^N-1$ have to be considered in the sum and thus strong connectivity is equivalent to strict positivity of all matrix entries in the \textit{reachability matrix} $R_N$ defined as
\begin{equation}
R_N:=\sum_{k=0}^{2^N-1} C^k_N = I_N + C_N + C^2_N + \cdots + C^{2^N-1}_N
\end{equation}
characterizing the transitive closure of the digraph $D(C_N)$ associated to $C_N$. As an alternative, the adjacency matrix with self-loops $D_N:=I+C_N$ can be used; in that case strong connectivity of the digraph is equivalent to primitivity of $D_N$ defined as the existence of an index $K\in \mathbb{N}$ such that $D_N^K>0$ entry-wise.
\\\\
Due to the binomial expansion for the (commuting) matrices $C_N$ and $I$
\begin{equation}
D^k_N = (I_N+C_N)^k = \sum_{j=0}^k {{k}\choose{j}} C_N^j = I_N+kC_N+k(k-1)C_N^2/2 + \cdots + C_N^k
\end{equation}
such an exponent $K$ is guaranteed to exist whenever $C_N$ is irreducible and vice versa. On the whole a powerful equivalence diagram is established, as shown in Figure 10.
\begin{figure}[h!]
\centering
\begin{tikzpicture}[node distance=3cm, every node/.style={draw, rectangle, minimum width=2.5cm, minimum height=1cm, align=center}]
    \node (SC) {Strong Connectivity \\ of $D(C_N)$};
    \node[right=of SC] (SP) {Strict Positivity \\ of $R_N$};
    \node[below=of SP] (P) {Primitivity \\ of $D_N$};
    \node[below=of SC] (IR) {Irreducibility \\ of $C_N$};
    \draw[<->] (SC) -- (SP);
    \draw[<->] (SP) -- (P);
    \draw[<->] (P) -- (IR);
    \draw[<->] (IR) -- (SC);
\end{tikzpicture}
\caption{Equivalences for Graph and Matrix Properties}
\end{figure}
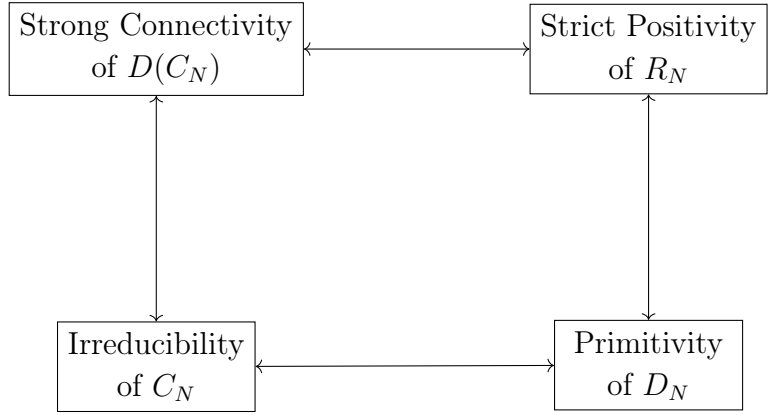
\\\\
This can now be exploited to prove that the connectivity is independent of lattice sizes in terms of $R_N>0$ entry-wise for all $N$. For compact notation in this proof, the summation/multiplication indices in
\begin{equation}
A_N = \sum_{i=1}^N \sigma_i^{-} = \sum_{i=1}^N I^{\otimes (i-1)}\otimes \sigma^- \otimes I^{\otimes(N-i)}
\end{equation}
 and
\begin{equation}
B_N = \prod_{i=1}^N \sigma_i^+ = \prod_{i=1}^N I^{\otimes (i-1)}\otimes \sigma^+ \otimes I^{\otimes(N-i)}
\end{equation}
will be suppressed; that way the deformed adjacency matrix is written as
\begin{equation}
C_N = \prod \sigma^{+} + \sum \sigma^{-} = B_N + A_N
\end{equation}
and
\begin{equation}A^N_N=N!\prod \sigma^-,\quad
A^{N+1}_N=0.
\end{equation}
The reachability matrix $R_N$ thus evaluates to 
\begin{equation}
R_N = \sum_{k=0}^{2^N-1} C^k_N = \sum_{k=0}^{2^N-1} \left[\sum \sigma^{-} + \prod \sigma^{+}\right]^k
\end{equation}
This expression is a binomial expansion for the noncommuting matrices $A_N$ and $B_N$; due to the nilpotency of both of them, the resulting sum includes many zero terms and reduces to
 \begin{equation}
 R_N = \sum_{k=1}^{2N}\sum_{j=1}^k  \left(\sum \sigma^-\right)^{k-j} \left(\prod \sigma^+\right) \left(\sum \sigma^-\right)^{j-1} +\sum_{k=1}^{N} \left(\sum \sigma^-\right)^k\end{equation}
Valid for any number of lattice sites, the formula yields for $R:=R_N$ 
\begin{equation}
R_{mn}= \sum_k \sum_j \left[ \left( \sum \sigma^{-} \right)_{m1}^{k-j} \left( \sum \sigma^{-} \right)^{j-1}_{2^Nn} \right] + \sum_k \left( \sum \sigma^{-} \right)_{mn}^k 
\end{equation}
as the entries of the reachability matrix in row $m$ and column $n$. Analyzing the powers of $A_N$, it is straightforward to see that there always exist choices for $j$ and $k$ to any indices $m$ and $n$ such that the respective matrix entry of $R$ is nonzero.
\\\\
This is easiest to conclude via verification for small system size $N=2$ and induction: the self-similarity of graphs and adjacency matrices encountered at each step $N \rightarrow N+1$ carries over to all powers $A_N^P$ for the very reason that those encode all paths of length $P$ of the underlying self-similar graph, so $A^P_N \rightarrow A^P_{N+1}$ displays a scaling behavior analogous to the one of $A_N \rightarrow A_{N+1}$ as a generalization of our previous observation: every step $A^p_N \rightarrow A^P_{N+1}$ corresponds to placing two copies of $A^P_N$ as the diagonal blocks of $A^P_{N+1}$ and a single copy of $A^{P-1}_N$ into the lower-left corner of $A^P_{N+1}$ for all powers $p\in\mathbb{N}$ and sizes $N\in\mathbb{N}$.
\\\\
Self-similarity patterns arise via iterations for the Lindbladian part of the adjacency matrices
\begin{equation}
A_{N+1}^P=
\left[
\begin{array}{
  >{\hspace{10pt}\rule{0pt}{15pt}}c<{\hspace{10pt}\rule{0pt}{15pt}}|
  >{\hspace{10pt}\rule{0pt}{15pt}}c<{\hspace{10pt}\rule{0pt}{15pt}}|
  >{\hspace{10pt}\rule{0pt}{15pt}}c<{\hspace{10pt}\rule{0pt}{15pt}}|
  >{\hspace{10pt}\rule{0pt}{15pt}}c<{\hspace{10pt}\rule{0pt}{15pt}}
  }
 & & & \\
A^{P}_{N-1} & 0 & 0 & 0 \\
 & & & \\
\hline
 & & & \\
A^{P-1}_{N-1} & A^{P}_{N-1} & 0 & 0 \\
 & & & \\
\hline
 & & & \\
A^{P-1}_{N-1} & 0 & A^{P}_{N-1} & 0 \\
 & & & \\
\hline
 & & & \\
A^{P-2}_{N-1} & A^{P-1}_{N-1} & A^{P-1}_{N-1} & A^{P}_{N-1} \\
 & & & \\
\end{array}
\right]
\end{equation}
as well as the Hamiltonian part
\begin{equation}
E_{N+1}^P=
\left[
\begin{array}{
  >{\hspace{10pt}\rule{0pt}{15pt}}c<{\hspace{10pt}\rule{0pt}{15pt}}|
  >{\hspace{10pt}\rule{0pt}{15pt}}c<{\hspace{10pt}\rule{0pt}{15pt}}|
  >{\hspace{10pt}\rule{0pt}{15pt}}c<{\hspace{10pt}\rule{0pt}{15pt}}|
  >{\hspace{10pt}\rule{0pt}{15pt}}c<{\hspace{10pt}\rule{0pt}{15pt}}
  }
 & & & \\
E^{P}_{N-1} & F(G_{N-1}) & F(G_{N-1}) & H(G_{N-1}) \\
 & & & \\
\hline
 & & & \\
F(G_{N-1})^T & E^{P}_{N-1} & E^{P-1}_{N-1} & F(G_{N-1}) \\
 & & & \\
\hline
 & & & \\
F(G_{N-1})^T & E^{P-1}_{N-1} & E^{P}_{N-1} & F(G_{N-1}) \\
 & & & \\
\hline
 & & & \\
H(G_{N-1})^T & F(G_{N-1})^T & F(G_{N-1})^T & E^{P}_{N-1} \\
 & & & \\
\end{array}
\right]
\end{equation}
with $F$ and $H$ matrix polynomials depending on $P$ and involving $G_{N-1}:=I_{N-1} + A_{N-1}$ for the recursions.
\begin{figure}[h!]
\includegraphics{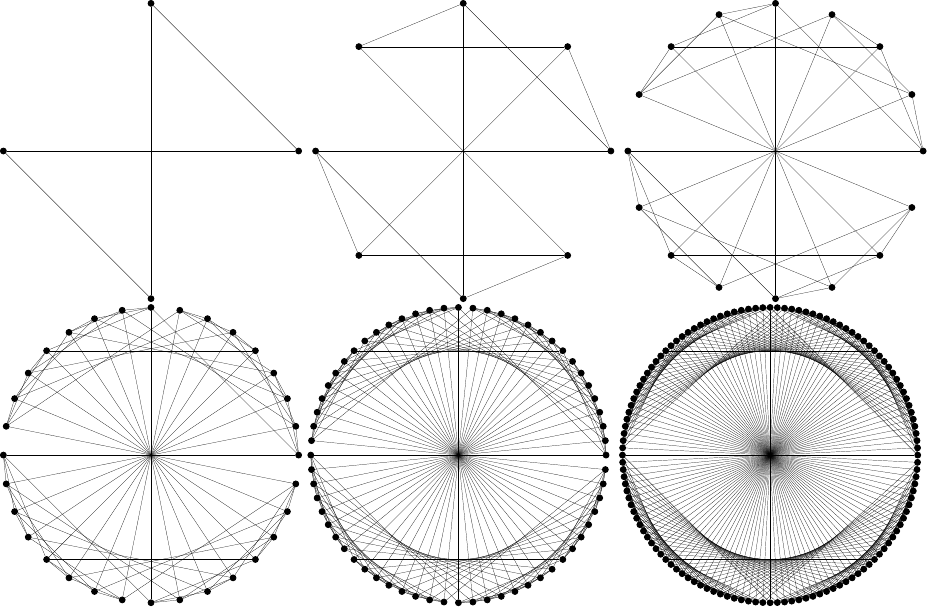}
\caption{Web Evolution for $\{L\}$}
\end{figure}
\begin{figure}[h!]
\centering
%
\caption{Limit Web for $\{L\}$}
\end{figure}
\begin{figure}[h!]
\includegraphics{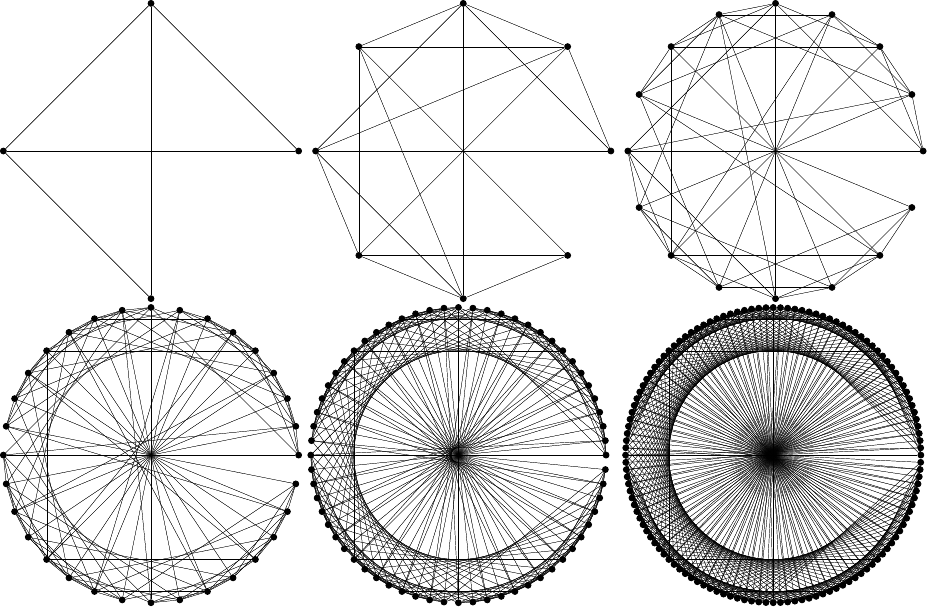}
\caption{Web Evolution for $H-i\sum L^*L/2$}
\end{figure}
\begin{figure}[h!]
\centering
%
\caption{Limit Web for $H-i\sum L^*L/2$}
\end{figure}
\section{Conclusion}
Exact graph self-similarity in the thermodynamic limit $N \rightarrow \infty$ has thus been shown; one crucial ingredient in the proof is the exact type of Jordan matrices $J$ that arise from the set of generators: the algebraic graph theory method works whenever a $J$-cyclic subspace equals the full space.
\\\\
This condition characterizes \textit{nonderogatory} Jordan matrices $J$: the existence of a cyclic vector $c \in V$ for $J$ is defined as the property of the set $\{J^{p}c\,|\,p\in \{0,...,\dim(V)-1\}\}$ spanning the whole space $V$. The important subspaces for our purposes are therefore both the invariant and cyclic subspaces of the generators in Jordan form.
\\\\
Whenever the given generating set in accordance with the Yoshida criterion contains a nonderogatory Jordan matrix in its entire linear span, it ensures the generation of the full matrix algebra in case of strong connectivity. Nonderogatory matrices have the defining property that their minimal and characteristic polynomials coincide and thus all geometric multiplicities equal one; in the generalized eigenbasis this is equivalent to the existence of exactly one Jordan block for each element of the spectrum.
\\\\
Important is thus the interplay between cyclic and invariant subspaces for the consequences entailed by nonderogatory generators in their generalized eigenbasis: whenever one of the operators in the span of $\{H-i\sum L^*L/2, L\}$ is similar to a nonderogatory Jordan matrix, strong connectivity of the associated digraph guarantees the generation of the full algebra.
\\\\
The simplest example for our model is the case of one spin site with rescaled jump $L/\sqrt{\gamma} =: L = \sigma^{-}=|0\rangle\!\langle 1|$ already being in Jordan normal form. This matrix is nonderogatory for the very reason that it is just a single Jordan block, so there exists an element $c \in \mathbb{C}^2$ cyclic for $L$ (i.e. a state $c$ yielding $\mathbb{C}^2$ as an $L$-cyclic subspace). The cyclic $c$ is found as $|1\rangle$ and generates the whole space through the repeated action of $L$ to yield the complete basis: $I|1\rangle = |1\rangle,L|1\rangle=|0\rangle , L^2|1\rangle=0$.
\section{Acknowledgements}
The authors are grateful for support by a research grant (42085) from VILLUM FONDEN. BB acknowledges funding by the French National Research Agency (ANR) under project ANR-24-CPJ1-0150-01.
\section{References}
\bibliographystyle{iopart-num}
\bibliography{new3}
\end{document}